\documentclass[prb,twocolumn,showpacs,preprintnumbers,amsmath,amssymb,floatfix]{revtex4}
\pdfoutput=1

\usepackage{graphicx}
\usepackage{dcolumn}
\usepackage{bm}
\usepackage{color}
\usepackage{amsmath}
\usepackage{multirow}

\newcommand{\MMoSe}{$M_2$Mo$_6$Se$_6$}
\newcommand{\T}{Tl$_2$Mo$_6$Se$_6$}
\newcommand{\I}{In$_2$Mo$_6$Se$_6$}
\newcommand{\R}{Rb$_2$Mo$_6$Se$_6$}
\newcommand{\C}{Cs$_2$Mo$_6$Se$_6$}

\usepackage{flafter}

\begin{document}

\title{\boldmath Phonon Mode Spectroscopy, Electron-Phonon Coupling and the Metal-Insulator Transition in Quasi-One-Dimensional {\MMoSe}}

\author{A.P. Petrovi\'c}
 \email{appetrovic@ntu.edu.sg}
\author{R. Lortz}
\author{G. Santi}
\author{M. Decroux}
\author{H. Monnard}
\author{\O. Fischer}
\affiliation{DPMC-MaNEP, Universit\'e de Gen\`eve, Quai Ernest-Ansermet 24, 1211 Gen\`eve 4, Switzerland}
\author{L. Boeri}
\author{O.K. Andersen}
\affiliation{Max Planck Institute for Solid State Research, Heisenbergstrasse 1, D-70569 Stuttgart, Germany}
\author{J. Kortus}
\affiliation{Institut f\"{u}r Theoretische Physik, TU Bergakademie Freiberg, Leipziger Strasse 23, D-09596 Freiberg, Germany}
\author{D. Salloum}
\author{P. Gougeon}
\author{M. Potel}
\affiliation{Sciences Chimiques, CSM UMR CNRS 6226, Universit\'e de Rennes 1, Avenue du G\'en\'eral Leclerc, 35042 Rennes Cedex, France}

\date{\today}

\begin{abstract}
We present electronic structure calculations, electrical resistivity data and the first specific heat measurements in the normal and superconducting states of quasi-one-dimensional {$M_2$Mo$_6$Se$_6$} ($M$ = Tl, In, Rb).  {Rb$_2$Mo$_6$Se$_6$} undergoes a metal-insulator transition at $\sim$ 170~K: electronic structure calculations indicate that this is likely to be driven by the formation of a dynamical charge density wave.  However, {Tl$_2$Mo$_6$Se$_6$} and {In$_2$Mo$_6$Se$_6$} remain metallic down to low temperature, with superconducting transitions at $T_{c}$ = 4.2~K and 2.85~K respectively.  The absence of any metal-insulator transition in these materials is due to a larger in-plane bandwidth, leading to increased inter-chain hopping which suppresses the density wave instability.  Electronic heat capacity data for the superconducting compounds reveal an exceptionally low density of states $D_{E_{F}}$ = 0.055 states eV$^{-1}$ atom$^{-1}$, with BCS fits showing $2\Delta/k_{B}T_{c} \geq$ 5 for {Tl$_2$Mo$_6$Se$_6$} and 3.5 for {In$_2$Mo$_6$Se$_6$}.  Modelling the lattice specific heat with a set of Einstein modes, we obtain the approximate phonon density of states $F(\omega)$.  Deconvolving the resistivity for the two superconductors then yields their electron-phonon transport coupling function $\alpha_{tr}^{2}F(\omega)$.  In {Tl$_2$Mo$_6$Se$_6$} and {In$_2$Mo$_6$Se$_6$}, $F(\omega)$ is dominated by an optical ``guest ion'' mode at $\sim$ 5~meV and a set of acoustic modes from $\sim$ 10-30~meV.  {Rb$_2$Mo$_6$Se$_6$} exhibits a similar spectrum; however, the optical phonon has a lower intensity and is shifted to $\sim$ 8~meV.  Electrons in {Tl$_2$Mo$_6$Se$_6$} couple strongly to both sets of modes, whereas {In$_2$Mo$_6$Se$_6$} only displays significant coupling in the 10-18~meV range.  Although pairing is clearly not mediated by the guest ion phonon, we believe it has a beneficial effect on superconductivity in {Tl$_2$Mo$_6$Se$_6$}, given its extraordinarily large coupling strength and higher $T_c$ compared to {In$_2$Mo$_6$Se$_6$}.   
\end{abstract}


\maketitle

\section{INTRODUCTION}

The {\MMoSe} ($M$ = Tl, In, Rb, Li, Na, K, Cs) system was first discovered by Potel {\itshape{et al.}} \cite{Potel-1980} and is closely related to the well-known quasi-three-dimensional (quasi-3D) Chevrel Phase compounds.  Rather than comprising individual ``zero-dimensional'' Mo$_6$$X_8$ ($X$ = S, Se, Te) octahedral clusters coupled by a metallic cation, these materials are composed of quasi-1D (Mo$_6$Se$_6$)$_\infty$ chains oriented along the {\itshape{z}} axis, weakly coupled by $M$ ions.  Only {\T} and {\I} are superconducting, with T$_c$ = 3 - 6.5~K (varying between samples) and $\sim$ 2.9~K respectively.  In contrast, {\R} undergoes a broad metal-insulator transition between 100~K and 200~K.~\cite{Tarascon-1984}  Little data currently exists in the literature for the remaining members of the family, although it is known that they exhibit similar metal-insulator transitions and do not become superconducting under ambient pressure at low temperature.~\cite{Hor-1985,Hor-1985-2}  

Reduced dimensionality and its effect on superconductivity remains one of the central issues in contemporary condensed matter physics research.  Since the late 1970s, numerous unconventional superconductors displaying highly anisotropic properties in both the normal and superconducting states have been discovered.  Among these, notable examples include the quasi-2D high-temperature cuprate superconductors (HTS) and the quasi-1D organic Bechgaard salts.  However, the most strongly 1D superconductors synthesized to date have attracted remarkably little attention over the years.  {\T} and {\I} boast anisotropy ratios $\epsilon=H_{c2}^{\parallel}/H_{c2}^{\perp} \geq$ 12.0 and 17.2 respectively,~\cite{Petrovic-2007,Armici-1980,Geserich-1986} significantly greater than $\epsilon \approx$ 8.5 in (TMTSF)$_2$ClO$_4$.~\cite{Lee-2002}  Furthermore, these materials do not possess any intrinsic magnetism, thus rendering them an ideal uncomplicated system for the study of low-dimensional superconductivity.  

In comparison with the HTS, the number of publications existing for {\MMoSe} is around three orders of magnitude smaller.  Early work concentrated on the electrical transport~\cite{Armici-1980,Lepetit-1984} and magnetic properties,~\cite{Brusetti-1988} immediately revealing large anisotropies in both the normal-state resistivity and superconducting coherence length for {\T}.  Two distinct classes of {\T} were identified~\cite{Armici-1980} by the behaviour of their longitudinal resistivity: $A$-type samples with conventional metallic behaviour down to low temperature or $B$-type samples displaying a broad minimum for $T<$ $\sim$ 80~K followed by an upturn reminiscent of charge density wave (CDW) formation.  More recent measurements~\cite{Petrovic-2007} have shown that the coherence length perpendicular to the chain axis $\xi_{\perp}$ is at most 75~{\AA}, a value not significantly larger than that found in some HTS.  In an important parallel with organic quasi-1D superconductors,~\cite{Lee-1997} the upper critical field perpendicular to the chain axis {\itshape{z}}, $H_{c2}^{\perp}$, does not saturate down to 50~mK.~\cite{Brusetti-1994}  Hall effect measurements by the same authors display a regime crossover at $T \approx$ 80~K which they attribute to the onset of a CDW or spin density wave (SDW).  However, there is no support for the formation of a CDW in normal-state resistivity curves for the $A$-type samples, which nonetheless display a Hall effect crossover.  Furthermore, the weak temperature-invariant diamagnetism in {\T} revealed by ac susceptibility measurements~\cite{Tarascon-1984} does not encourage a SDW interpretation.  

The discovery that the application of uniaxial stress along the {\itshape{z}} axis in {\T} suppresses superconductivity and induces a metal-insulator transition increased the evidence for this system being close to a CDW instability, in particular due to the non-linear $I-V$ curves and broadband noise characteristic of density wave motion observed.~\cite{Tessema-1991} Conversely, hydrostatic pressure increases the conductivity of {\T} in the normal state but still suppresses superconductivity.~\cite{Huang-1983}  

At first glance, the quasi-one-dimensional nature of the {\MMoSe} family renders them strong candidates to undergo a Peierls transition, so it was initially a mystery as to why {\T} and {\I} remained metallic down to low temperature.  Early attempts to resolve this question focussed on the calculated band structure~\cite{Kelly,Nohl} which displayed three contributions to the Fermi surface: a broad singly-degenerate quasi-1D Mo $d$ ``helix'' band and two doubly-degenerate 3D ``octahedral'' electron pockets at the zone boundary.  The occupancy of these 3D pockets was believed to stabilise the structure against a Peierls transition and the authors of all ensuing publications attempted to interpret their results within this multi-band framework.  However, band structure calculation techniques have significantly advanced since the non-self-consistent approaches of the early 1980s.  It is therefore instructive to recalculate the band structure of {\MMoSe} using a fully self-consistent method and compare our results with the existing calculations for {\T}.

The nature of the superconducting state in {\T} and {\I} also remains very unclear, particularly since recent scanning tunnelling microscopy (STM) experiments~\cite{Dubois-2007%,Petrovic-2010
} on the related quasi-3D Chevrel Phase superconductor PbMo$_6$S$_8$ have provided strong evidence for a highly anisotropic or noded %or multi-band 
gap function.  Inelastic neutron scattering measurements of the phonon density of states (PDoS)~\cite{Brusetti-1990} revealed a strong low-energy Einstein-like optical mode attributed to vibrations of the $M$ ``guest'' ions between the Mo$_{6}$Se$_{6}$ chains, as well as higher energy intra-chain modes similar to those seen in the 3D Chevrel phases.~\cite{Bader}  This finding is supported by early normal-state specific heat data~\cite{Bonjour-1985} which unfortunately lacked sufficient resolution to provide any information on the superconducting state.  However, no studies of the electron-phonon coupling were performed; nor have any tunnelling experiments been carried out on the {\MMoSe} system.  

Motivated by recent discoveries in boride~\cite{Lortz-2005,Lortz-2006} and $\beta$-pyrochlore~\cite{Hiroi-2007} systems where superconductivity is mediated by a low-energy rattling phonon, we therefore decided to measure the PDoS and electron-phonon coupling for {\T}, {\I} and {\R} by deconvolving normal-state specific heat and resistivity data.  These measurements are analysed in parallel with our specific heat and resistivity data below $T_{c}$ for $M$ = Tl,In and in the insulating phase for $M$ = Rb.  In addition, we have performed a complete theoretical analysis of the electronic structure and parameters governing the metal-insulator transition in {\MMoSe}.  By combining our experimental data with the trends indicated by our new band structures, we hope to remove some of the confusion surrounding the mechanism for superconductivity and the effects of low-dimensionality in this fascinating system on the borderline between superconducting and insulating instabilities.  

\section{THEORY}
\subsection{Crystal Structure}

\begin{figure}[htb]
\includegraphics[width=8.0cm,clip]{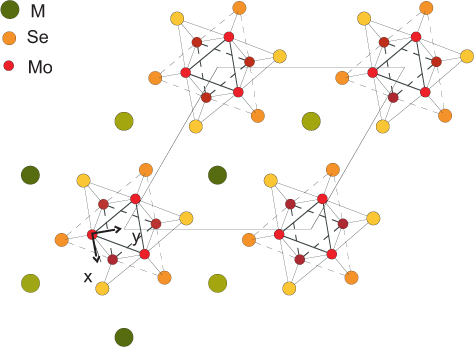}
\includegraphics[width=4cm,clip]{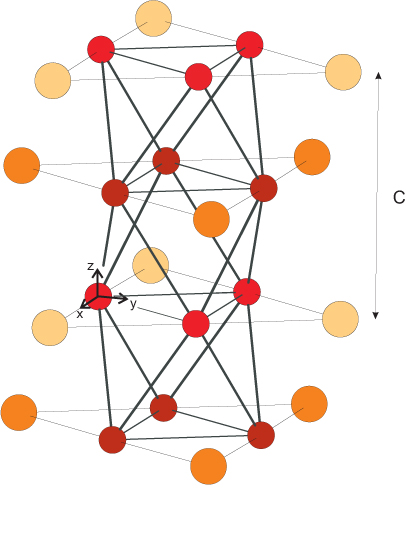}
\caption{\label{fig1}
Crystal structure of {\MMoSe} compounds in the 001 plane (above) and side view (below) of the (Mo$_6$Se$_6$)$_{\infty}$ chains. Darker (lighter) symbols indicate atoms sitting on even (odd) planes  respectively.  The local coordinate system used to plot the partial $Mo$, $Se$ characters in Fig.~\ref{fig2} is also shown.
}
\end{figure}

The crystal structure of the compounds with chemical formula {\MMoSe} is shown in Fig.~\ref{fig1}, viewed both parallel and perpendicular to the $z$ axis.  Mo and Se atoms form quasi-1D (Mo$_6$Se$_6$)$_\infty$ chains oriented along the {\itshape{z}} axis, separated by $M$ ions in a zig-zag formation.  The chains consist of inner Mo and outer Se triangles, stacked with a $c$/2 separation along the $z$ axis and rotated 180$^{\circ}$ with respect to each other.  The axes of the Mo-Se triangles are aligned with each other, but rotated by 10$^{\circ}$ from the lattice vectors in the $x-y$ plane.  We may consider the chains to be a linear condensation of Mo$_6$S$_8$ clusters via face-sharing of the Mo$_6$ octahedra.  These clusters are the building blocks of the related quasi-3D Chevrel phases.  

\begin{table}[htbp]
\begin{ruledtabular}
\begin{tabular}{l|cccc}
$M$           & $x_{Mo}$    & $y_{Mo}$ &$x_{Se}$ &   $y_{Se}$     \\
\hline
In          & 0.189      & 0.156   & 0.068  &   0.369       \\
Tl          & 0.187      & 0.154   & 0.067  &   0.366       \\
Rb          & 0.181      & 0.149   & 0.064  &   0.355       \\ 
\end{tabular}
\end{ruledtabular}
\caption{Optimized internal coordinates (Wyckoff positions) 
for Mo and Se in the
three {\MMoSe} compounds considered in this work ($M$=Tl,In,Rb).
\label{table2}}
\end{table}

The conventional unit cell is hexagonal (space group P6$_3$/$m$), and contains two formula units ({\em f.u.}); $M$ atoms occupy $2d$ positions while Mo and Se atoms occupy $6h$ positions.  The hexagonal lattice parameters (determined by X-ray diffraction) are $a_{H} =$ 8.854, 8.934, 9.257~{\AA} and $c_{H} =$ 4.493, 4.494, and 4.487~{\AA} respectively for In, Tl and Rb-based crystals.  It can immediately be seen that the inter-chain distance correlates with the atomic radius of the $M$ atom, whereas $c_{H}$ (and hence the intra-chain atomic separation) remains roughly constant regardless of $M$.  Furthermore, the intra-chain atomic separations are much smaller than the inter-chain distances.  For example, in Tl the shortest Mo-Mo distance (for two Mo atoms in the same triangle) is 2.66~{\AA}; the shortest Mo-Se distance is 2.695~{\AA} and the Se-Se separation is 3.767~{\AA}.  In contrast, the Mo-Mo inter-chain separation is 6.34~{\AA}.  

The crystal structure thus provides an immediate indication of the strong anisotropy present in this family of materials. In the following, we will show that a large anisotropy is also found in the electronic structure and discuss its consequences for the physical properties of {\MMoSe} compounds.

\subsection{Electronic Structure}

We have performed  $ab-initio$ Density Functional Theory (DFT) calculations of the electronic properties of {\MMoSe}, for $M$ =Tl, In, Rb, employing the the full-potential Linear Augmented Plane Wave (LAPW) method.~\cite{Andersen-1975,WIEN2k}  For all systems, we used the experimental lattice parameters and optimized the internal coordinates: these optimized values are given in table \ref{table2}.

For {\T}, where the experimental Wyckoff positions are known,~\cite{Potel-1980-2} the optimized coordinates agree with the experimental ones to better than $1~\%$; for In$_2$Mo$_6$Se$_6$ and Rb$_2$Mo$_6$Se$_6$ we could not compare with experimental data.  Structural optimization yielded intra-chain distances which do not depend (within computational accuracy) on the nature of the $M$ atom, whereas interchain distances increase by 5 $\%$  going from Tl,In to Rb due to the larger in-plane lattice constant.

\begin{figure}[htbp]
\includegraphics*[width=8.5cm]{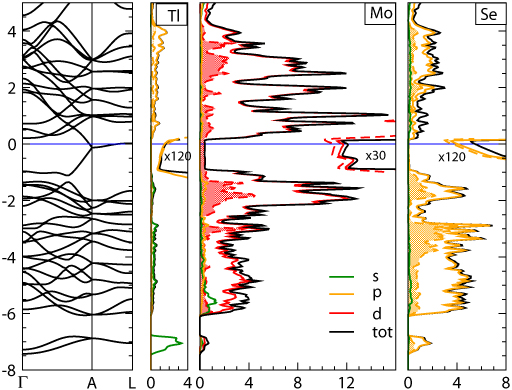}
\caption{\label{fig2} Band structure and partial DoSs  of {\T}; energies are in eV and measured w.r.t.~~the Fermi level, DoS are in states/(eV cell).  Colors indicate different $l$ characters (green=$s$, orange=$p$, 
red=$d$).  Shaded areas, delimited by dashed lines, refer to the partial characters (Tl $p_z$, Mo $d_{xz}$, Se $p_{x}$) which provide the largest contribution at E$_F$. For a definition of the axes and atoms, see Fig.~\ref{fig1}.  An enlargement of the DoS around the Fermi level is also shown; the numbers indicate the enlargement factor.
}
\end{figure}

Since the major features of the electronic structure of {\MMoSe} are the same for $M$=Tl, In, Rb, we will first present the electronic structure of Tl$_2$Mo$_6$Se$_6$ and then discuss the differences with In and Rb.  Our calculated electronic structure  reproduces the main features of the earlier non-self-consistent calculations,~\cite{Kelly,Nohl} except for one crucial difference in the position of the Fermi level.  The electronic bands and partial densities of states (DoS) are shown in Fig.~\ref{fig2}, with all energies measured relative to the Fermi level.  Se $s$ and Tl $d$ states form weakly-dispersing bands, located 15~eV and 12~eV below the Fermi level respectively.  The Tl 6$s$ states lie approximately 7~eV below the Fermi level; Tl is therefore in a nominal +1 valence state, effectively behaving like an alkali metal.  The bands lying between -5 and +5~eV are mainly of Mo $d$ and Se $p$ character, although there is a significant hybridization with Tl $p$ states as can be seen in Fig.~\ref{fig2}.  The Mo $sp$ bands are displaced to higher energies by covalent hybridisation with Se $p$ bands and hence lie above +5~eV.

\begin{figure}[h!tbp]
\includegraphics*[width=8cm]{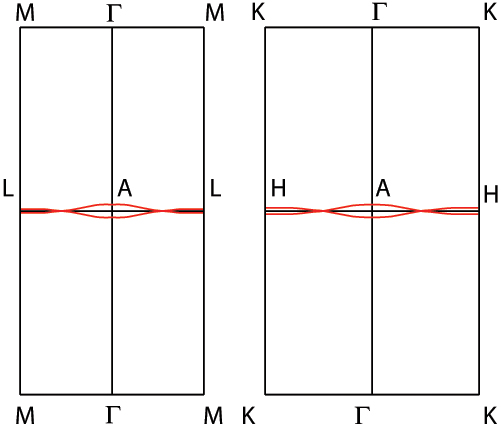}
\caption{\label{fig3}
Fermi surface of {\T}, in a vertical plane cutting through the center of the hexagonal Brillouin zone, shown in Fig.~\ref{fig1}. }
\end{figure}

\begin{figure}
\includegraphics*[height=4cm]{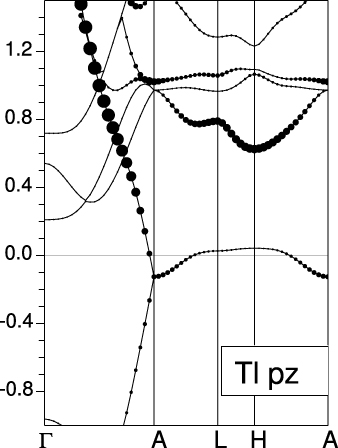}
\includegraphics*[height=4cm]{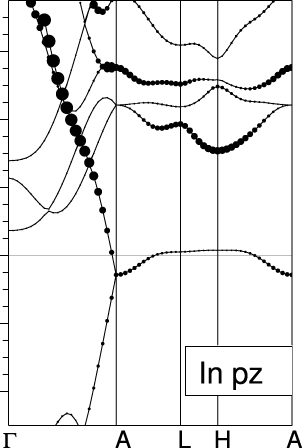}
\includegraphics*[height=4cm]{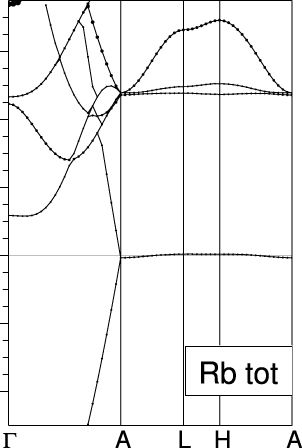}
\caption{\label{fig4}
Band structures of {\MMoSe}, for $M$=Tl,In and Rb decorated with partial $M$ characters. Energies, measured w.r.t. the Fermi level, are in eV.}
\end{figure}

The 18 Se $p$ bands are centered around -4~eV and overlap with the lower portion of the 30 Mo $d$ bands, which extend over $\pm$5~eV around the Fermi level.  These are divided into 12 bonding and 16 antibonding states, separated by a pseudogap $\sim$~1~eV wide around the Fermi level.  Two highly 1D bands cross the gap: they derive from the zone-folding of a single helix band. %made of the  two Bloch states which wind up along the chain, %in a lefthand and a righthand helices.
The Fermi level cuts the band structure at $k_z$=$\pi/c$, exactly at half filling.  The corresponding Fermi Surface is shown in Fig.~\ref{fig3}.  In contrast with earlier calculations, we do not find the additional small 3D pockets at the Fermi level which were previously alleged to be responsible for the stability of the chains against any Peierls distortion and ensuing density wave transition.~\cite{Kelly}  The octahedron band (of Mo $d$ $3z^2-r^2$ and $x^2-y^2$ character) responsible for these pockets in fact lies $\sim$~0.1~eV above E$_F$.  Fig.~\ref{fig4} shows that {\I} and {\R} have a very similar band structure to  {\T}, i.e. there are no other bands at $E_F$ except for the helix band.  Small variations in the dispersion of this band must therefore control the stability of the {\MMoSe} family.

Let us now consider the relative anisotropy of the {\MMoSe} series, in terms of details of the dispersion of the helix band, using a simplified tight-binding model with large out-of-plane ($W$) and small in-plane ($w$) bandwidths.  In the three rightmost panels of Fig.~\ref{fig2}, we highlight the  partial orbital characters which give the largest contribution to the helix band: the dominant contribution is from Mo d$_{xz}$ states.  (The labels of the orbitals refer to the local coordinate systems
centered on the Mo atoms, shown in Fig.~\ref{fig1}).

\begin{table}[tbh]
\begin{ruledtabular}
\begin{tabular}{l|ccc}
                        & Tl          & In          & Rb         \\
\hline
$W$                     & 7           & 7           & 8          \\
$w$                     & 0.17        & 0.14        & 0.02       \\
$\lambda_c  $           & 0.11        & 0.11        & 0.07       \\
$D$                     & 6.74        & 6.74        & 7.70       \\
$\omega_{0,1}$          & 22          & 22          & 22         \\
$\omega_{0,2}$          & 27          & 27          & 27         \\
$\lambda(\omega_{0,1})$          & 0.10        & 0.10        & 0.13       \\
$\lambda(\omega_{0,2})$          & 0.07        & 0.07        & 0.09       \\
$\delta_0$                       &  0          &  0          & 0.17       \\
$\omega_{1}$                     &6.6 & 6.6 & 11.2\\
$\omega_{2}$                     &8.2 & 8.2 & 11.4\\
\end{tabular}
\end{ruledtabular}
\caption{Parameters governing the metal-insulator transition in {\MMoSe} ($M$~=~Tl,In,Rb) from local density approximation (LDA) calculations.  $W$ and $w$ are the out-of-plane and in-plane bandwidths of the helix band in eV: they determine the dimensionless critical coupling constant $\lambda_c$ (see Eq.~\ref{eq:lambdac}).  $\omega_{0,i}$ are the bare frequencies of the two Peierls modes in meV. $\lambda(\omega_{0,i})$ are the corresponding dimensionless bare electron-phonon coupling constants (see Eq.~\ref{eq:lambda0}); 2 $\delta_0$ is the value of the Peierls gap (Eq.~\ref{eq:delta0}) in eV.  The renormalized phonon frequencies $\omega_{i}$ (in meV) are obtained from Eqs.~\ref{eq:omegah} and ~\ref{eq:omegaoff} for Tl, In and Rb respectively.
\label{table0}}
\end{table}

In fact, the Bloch states that  form this band are built from the in-phase linear combination of the three equivalent Mo $xz$ orbitals which sit on each Mo$_3$ triangle.  Due to the presence of a two-fold screw axis along the center of the chain, they form right and left-handed helices which wind up along the chain axis. (For a definition of the axes, see Fig.~\ref{fig1}).  The out-of-plane bandwidth $W$ is thus given by the hopping of one Mo d$_{xz}$ to another Mo d$_{xz}$ orbital on the next plane; the in-plane bandwidth is given by the hopping between the chains, mediated by Se $p_x$ and Tl(In) $p_z$ or Rb $s$ orbitals (illustrated by yellow and green lines respectively in Fig.~\ref{fig2}).

Due to symmetry, hopping via Tl(In) $p_z$ orbitals is more effective than via Rb $s$ orbitals.  In Fig.~\ref{fig4} we show a close-up of the band structures of {\MMoSe}, decorated with the partial characters associated with the $M$ atoms.  Here, we observe that the contribution of Tl and In $p_z$ states to the helix band is much stronger than that of the Rb $s$ states.  This agrees well with the significant In $p$ orbital contribution to the conduction band in {\I} suggested by nuclear magnetic resonance spectroscopy.~\cite{Chew-1994}  Correspondingly, the in-plane bandwidth is reduced by a factor of $\sim 10$ in {\R}.  From a fit of these three band structures, we extract the values of $W$ and $w$ reported in table~\ref{table0}.

\subsection{Mechanisms for Metal-Insulator Transitions}

The first possibility for the metal-insulator transition is a CDW instability, due to the interaction of the helix band with phonons (i.e.~~a Peierls transition). In {\MMoSe}, there are two Peierls-like modes which could lead to dimerization of the chains and open a gap at $E_F$: the eigenvectors are such that the Mo triangular units on subsequent planes move out-of-phase with respect to each other and the surrounding Se triangles can then be displaced either in or out of phase.

Examining these two modes, we use local density approximations (LDA) to determine two bare frequencies $\omega_{0,1}$ and  $\omega_{0,2}$ which we have found to be independent of the nature of the $M$ guest ion.  Our calculated values are $\omega_{0,1}=22$ meV and $\omega_{0,2}=27$ meV.  These frequencies include the response of the whole electronic band structure to the phonon perturbations, with the exception of the helix band, which we calculate analytically below.  This procedure is more accurate than a direct LDA calculation of the phonon frequency, which becomes extremely inaccurate as the in-plane bandwidth $w$ of the helix band tends to zero.  

The effect of both modes on the band structure is the same: a frozen displacement $e u$ of the atoms along the renormalized eigenvector opens a gap $2 \delta=2 D u$ at the Fermi wavevector $k_z=\pi/c$.  This defines the deformation potential $D$ which (together with the bare phonon frequencies  $\omega_{0,i}$ and the DoS at the Fermi level $N(0)=1/W$) determines the bare electron phonon coupling constants  $\lambda(\omega_{0,i})$ through the relation:
\begin{equation}
\label{eq:lambda0}
\lambda(\omega_{0,i})=\frac{D^2}{ W M \omega_{0,i}^2}.
\end{equation}
Due to the finite in-plane bandwidth of the helix band, a Peierls transition can take place only if this bare electron-phonon coupling constant $\lambda(\omega_{0,i})$ exceeds a critical value $\lambda_c$: 

\begin{equation} 
\label{eq:lambdac}
\lambda_c=\frac{1}{2\left(\ln(W/w)+1\right)}.
\end{equation}

If $\lambda(\omega_{0,i}) < \lambda_c$, there is no Peierls transition, but the bare phonon frequency is renormalized:

\begin{equation}
\label{eq:omegah}
\omega_i=\omega_{0,1}\sqrt{\left(1-\frac{\lambda(\omega_{0,i})}{\lambda_c}\right)}
\end{equation}

If $\lambda(\omega_{0,i}) > \lambda_c$, there is a Peierls transition, with a gap $2 \delta_0=2 D u_0$:
\begin{equation}
\label{eq:delta0}
2 \delta_0= 2 W \exp (-\frac{1}{2 \lambda(\omega_{0,i})}),
\end{equation}

The frequency of the $i^{th}$ phonon in the off-centre position is then given by: 
\begin{equation}
\label{eq:omegaoff}
\omega_i=\omega_{0,i}\sqrt{2 \lambda(\omega_{0,i})}.
\end{equation}

Table \ref{table0} lists the relevant LDA parameters for $M$=Tl,In,Rb.  It may readily be seen that all {\MMoSe} compounds are close to the CDW instability, since the values of the critical interaction parameters $\lambda_c$ are very low compared to the bare electron-phonon coupling parameters $\lambda_{0,i}$.

For {\T} and {\I}, the bare electron-phonon coupling constants for both the lower and the higher Peierls modes are slightly below $\lambda_c$; for {\R} $\lambda(\omega_{0,1})$ is well above the critical value, thus indicating that a Peierls transition is likely to occur with a gap $2 \delta = 0.34$ eV.  However, due to the small value of  $\lambda(\omega_{0,1})$, the off-center minimum of the phonon potential is very shallow, implying that the transition is of dynamical rather than static character.  This means that the density wave formation may not be characterised by a static structural modulation, as is the case for ``classical'' CDW systems such as NbSe$_2$.   

An alternative explanation for the observed metal insulator transition in {\MMoSe} compounds could be a spin density wave (SDW) instability.  A staggered antiferromagnetic order of Mo$_3$ units is the spin analogue of the Peierls distortion.  The spin density wave (SDW) transition is regulated by the magnetic coupling constant $\lambda_I$=$\frac{1}{4}\frac{I}{3}\frac{1}{W}$, where the Stoner parameter $I$ can be estimated from the atomic value for Mo: 0.6 eV.  $\lambda_I$ is one order of magnitude smaller than $\lambda_c$ in Tl, In and Rb; we therefore do not find a stable antiferromagnetic (AFM) solution by LDA for any of the {\MMoSe} compounds.

However, AFM ordering is also favored by the Coulomb repulsion $U$, which is much larger than Hund's coupling $I$. Constrained LDA calculations give $U$=5.1 eV and $J$=0.62 eV per Mo atom in Tl$_2$Mo$_6$Se$_6$.  With these values, we estimate a magnetic coupling constant $\lambda_U$=0.12 for Tl,In and 0.11 for Rb, which implies that all {\MMoSe} are on the verge of a SDW transition due to strong electronic correlations. 

For both charge (Peierls) and spin (AFM) density wave transitions, the stronger tendency of {\R} towards insulating behaviour compared with  {\T} and {\I} is due to the smaller value of the critical coupling constant $\lambda_c$.  This is mainly caused by the reduced in-plane bandwidth of the helix band, which we recall is one order of magnitude smaller for $M$~=~Rb than for $M$~=~In, Tl.  Further details of our calculations and the analytical model employed will be discussed elsewhere.\cite{LB}  We will now turn to our experimental data, which confirms our experimental calculations and provides further insight towards the metal-insulator transition.

\section{EXPERIMENT}
\subsection{Sample Preparation and Experimental Details}

Needle-like crystals of dimensions approximately 4~mm$\times$300~$\mu$m$\times$100~$\mu$m and mass $\approx$ 800~$\mu$g were synthesised by different methods depending on the thermal stability of the compounds: {\T} and {\R} were prepared in sealed molybdenum crucibles at 1700~$^\circ$C and 1500~$^\circ$C respectively, whereas {\I} was prepared in an evacuated sealed silica tube at 1100~$^\circ$C.  Their crystalline structures were verified by the mono-crystal diffraction method using a KAPPA CCD NONIUS and exhibit a slight cation deficiency with occupancy factors 0.95, 0.94 and 0.93 for In, Tl and Rb based compounds respectively.  Larger polycrystalline samples of {\T} and {\I} of mass 10-75~mg were also produced by the same methods.  

\begin{figure}[h!]
\centering
\includegraphics [width=8.5cm,clip] {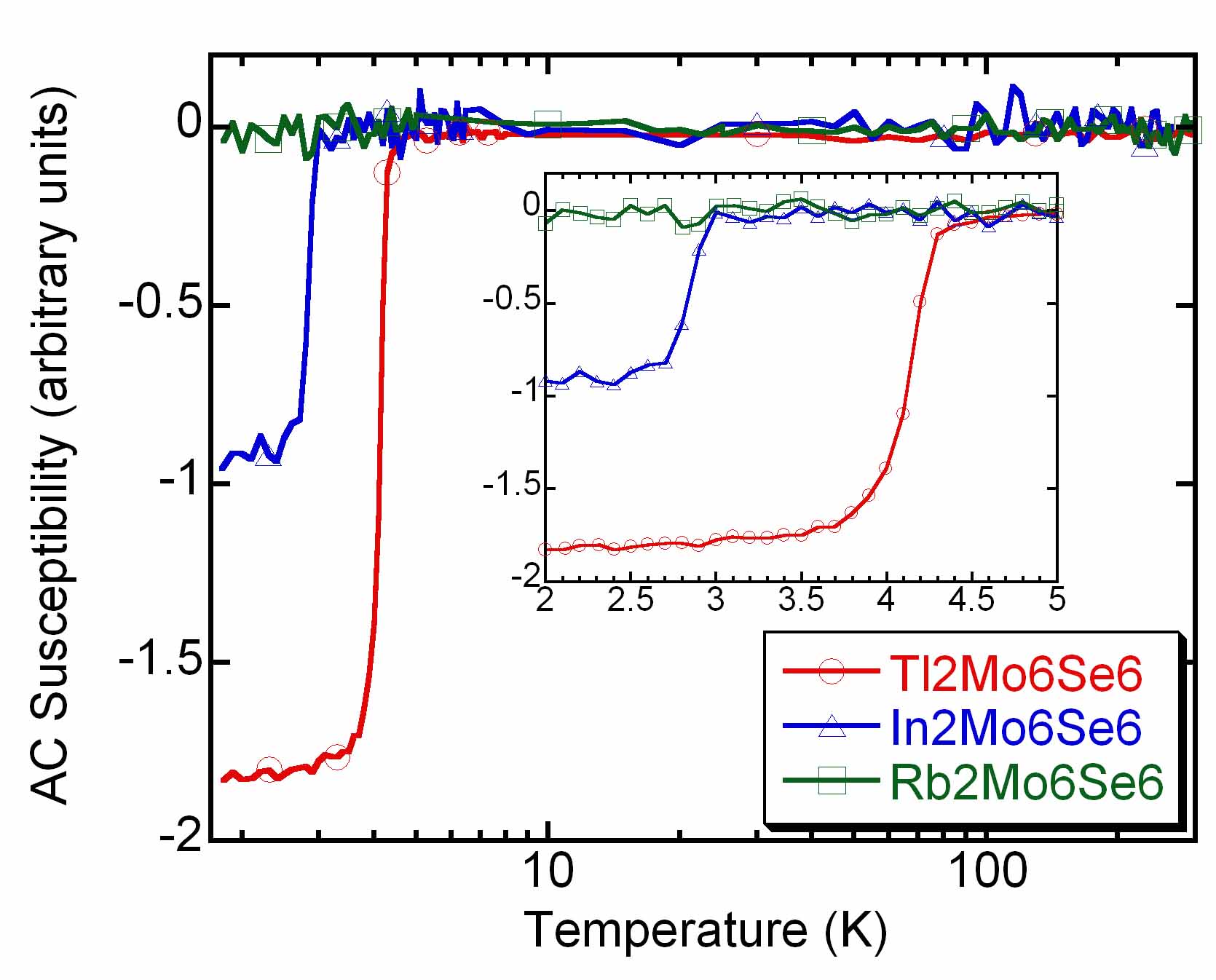}
\caption{\label{Fig_1} AC susceptibility of {\T}, {\I} and {\R} from 1.8-300~K.  Inset: zoom onto superconducting transitions, with $T_{c}$=4.2~K and 2.85~K in {\T} and {\I} respectively.  As is the case for all graphs in this work, only a small fraction (typically 10-20~{\%}) of the data-points measured are explicitly marked for clarity.}  
\end{figure}

The samples were initially characterized by AC Susceptibility, measured using a Quantum Design$^{\textrm{TM}}$ Physical Property Measurement System (QD PPMS).  As shown in Fig.~\ref{Fig_1}, {\T} and {\I} both exhibit superconducting transitions at $T_c$=4.2~K and 2.85~K, with ${\Delta}T_c$=0.7~K and 0.5~K respectively.  {\R} displays a constant weak diamagnetic signal down to the lowest temperature measured (1.7K) with no sign of superconductivity being observed.  No significant increase in transition width was observed for the polycrystals compared to the single crystals, indicating a high sample quality.  We stress that all members of the {\MMoSe} family remain stable under atmospheric conditions and none of our samples has exhibited any ageing effects.  

AC resistivity was measured using the same QD PPMS with a Helium-3 insert from 0.35-300~K.  Four gold contacts of thickness $\sim$5~nm were sputtered onto single crystals of each compound and 50~$\mu$m gold wires attached using silver epoxy glue.  This method yielded contact resistances ${\sim}1~{\Omega}$.  Short (1~s) pulses of a small AC current (0.02~mA, 470~Hz) were used to minimise any heating effects in the sample at low temperature.  

Specific heat was initially also measured in the QD PPMS using a standard relaxation technique from 0.35-300~K.  The largest homogeneous polycrystalline samples available for each compound were mounted using a measured quantity of Wakefield$^{\textrm{TM}}$ grease (whose contribution to the heat capacity was later subtracted).  However, due to the extremely small density of states at the Fermi level the QD PPMS was unable to detect the superconducting transition in {\T} and {\I}.  High-sensitivity relaxations from 1.3-10~K were therefore carried out on single crystals in our dedicated specific heat laboratory, enabling us to study the superconducting transition with a magnetic field both perpendicular and parallel to the {\itshape{z}} axis.  

\subsection{Characteristics of the Superconducting State in ${\textrm{Tl}}_{2}{\textrm{Mo}}_{6}{\textrm{Se}}_{6}$ and ${\textrm{In}}_{2}{\textrm{Mo}}_{6}{\textrm{Se}}_{6}$}

\begin{figure}[h!]
\centering
\includegraphics [width=8.5cm,clip] {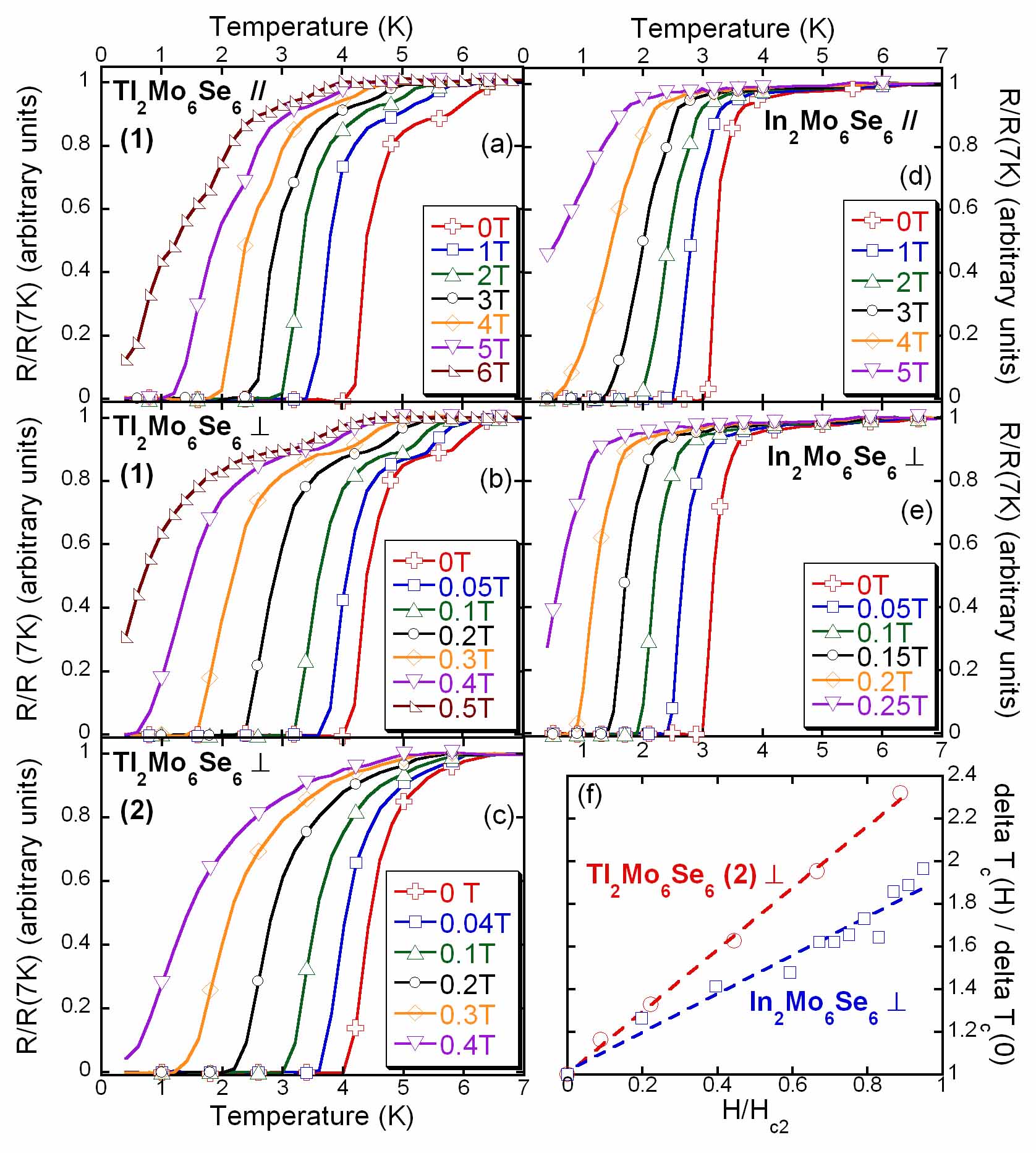}
\caption{\label{Fig_2}(a)-(e): Resistive transitions in {\T} and {\I} with field both parallel and perpendicular to the {\itshape{z}} axis.  Two samples of {\T} are shown: Sample (1) displaying a double transition indicating an inhomogeneous Tl content and Sample (2) with a single broad transition.  (f): Normalised transition widths ${\Delta}T_{c}(H)/{\Delta}T_{c}(0)$ as a function of normalised magnetic field $H/H_{c2}$.  For a transition in magnetic field $H$, ${\Delta}T_{c}$ is defined as the temperature difference between resistivities of 5\% and 95\% of the saturated normal-state value at 7~K.}
\end{figure}

As previously reported,~\cite{Petrovic-2007} resistive transitions into the superconducting state of {\T} exhibit an anomalous broadening under an applied magnetic field $H$, similar to that seen in the HTS.  We have performed similar measurements on {\I} and show these in Fig.~\ref{Fig_2}, together with the results in {\T} for comparison.  It can immediately be seen that {\I} also displays a broadening of the transition with increasing field, although the effect is less spectacular than in {\T}.  This is highlighted in Fig.~\ref{Fig_2}(f), where we have plotted the normalised transition width ${\Delta}T_{c}(H)/{\Delta}T_{c}(0)$ as a function of normalised perpendicular magnetic field $H/H_{c2}$ for both {\T} and {\I}.  Both materials display a linear behaviour in ${\Delta}T_{c}(H)/{\Delta}T_{c}(0)$ as the field increases, with the gradient for {\T} $\sim~30\%$ larger than that of {\I}.  

The width of a superconducting transition is governed by two parameters: the intrinsic homogeneity of the superconductor and the narrow thermal fluctuation-dominated critical region which surrounds any phase transition.  Inhomogeneity contributions are field-independent, but for 3D fluctuations the critical region, whose width is given by the Ginzburg parameter $G_{3D} = (k_{b}T_{c}/\sqrt{2}{\xi}_{\perp}^{2}{\xi}_{\parallel}H_{c}(0)^{2})^{2}$ multiplied by $T_{c}$, is weakly field-dependent, since $G_{3D}(H)~{\approx}~G_{3D}^{1/3}(H/H_{c2})^{2/3}$.  This clearly cannot explain our data, since we observe ${\Delta}T_{c}$ to increase linearly as $H$ increases.  However, Mishonov {\itshape{et al.}}~\cite{Mishonov-2003} derived a quasi-1D Ginzburg parameter for a superconducting nanowire: 

\begin{equation}
G_{1D} = \frac{k_{B}}{8\sqrt{\pi}{\Delta}C{\xi}(0)S}
\end{equation}

where ${\Delta}C$ is the specific heat jump at $T_{c}$, $S$ is the cross-sectional area of the nanowire and we only consider longitudinal fluctuations so ${\xi(0)}~{\equiv}~{\xi_{\parallel}(0)}$.  Modelling {\MMoSe} as a weakly-coupled assembly of superconducting filaments, each with the radius of a single Mo$_{6}$Se$_{6}$ chain, we calculate critical region widths $G_{1D}T_{c}$ = 1.5~K and 2.0~K for {\T} and {\I} respectively.  

\begin{figure}[htbp]
\centering
\includegraphics [width=8.5cm,clip] {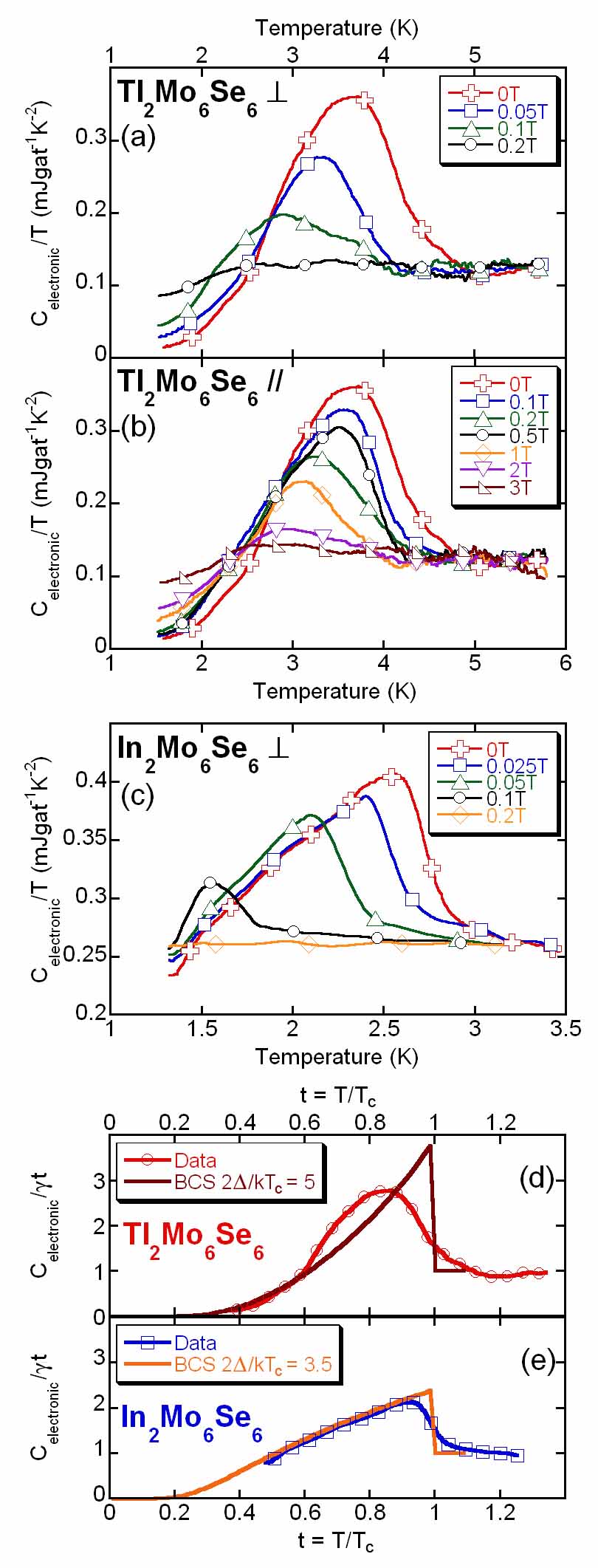}
\caption{\label{Fig_4}(a)-(c): Superconducting transitions in {\T} and {\I} seen by specific heat with field both parallel and perpendicular to the {\itshape{z}} axis.  (d),(e): BCS fits of the zero-field transition in {\T} and {\I}.}
\end{figure}

Similar trends are also seen in the electronic contribution to the heat capacity (Fig.~\ref{Fig_4}), with {\T} and {\I} behaving very differently in a magnetic field.  The onset temperature $T_{ons}$ of the specific heat jump in {\I} is rapidly and uniformly displaced to lower temperature with increasing $H$, whereas regardless of the applied field strength or orientation, $T_{ons}$ does not drop below $\sim$~4K in {\T}.  It should be noted that {\I} shows an abnormally large electronic contribution to the specific heat $\gamma$ in proportion to its specific heat jump, suggesting that only around 50$\%$ of the electrons at the Fermi level become superconducting.  This scenario could be explained by a slight variation in the sample stoichiometry leading to a CDW coexistent with superconducting regions in the same crystal, hence also supporting our observation that {\I} is more anisotropic than {\T}.  Correcting for this anomaly in {\I}, we estimate a Sommerfeld constant $\gamma \approx$ 0.13~mJgat$^{-1}$K$^{-2}$ for both {\T} and {\I}, corresponding to a dressed density of states at the Fermi level $D_{E_{F}}$ = 0.055 states eV$^{-1}$ atom$^{-1}$.

We have attempted to fit the low-temperature specific heat in {\T} and {\I} using a standard BCS s-wave single-band $\alpha$-model, as shown in Fig.~\ref{Fig_4}(d).  As discussed in the previous section, our band structure calculations indicate that only a single band (the 1D Mo $d$ helix) crosses the Fermi level, thus eliminating any possibility of multi-band superconductivity in {\MMoSe}.  Although the jump we measure at $T_{c}$ is thermally broadened in each case, this yields an excellent fit for {\I} with a gap value of 0.4~meV corresponding to 2$\Delta_{0}/k_{B}T_{c}$ = 3.5, just below the standard weak-coupling BCS value 3.52.   An unusual hump on the back of the peak in {\T} renders the fit more difficult in this compound, but by using entropy conservation it is still possible to estimate a gap $\Delta_{0} \geq$ 1.1~meV and 2$\Delta_{0}/k_{B}T_{c} \geq$ 5, placing it in the extreme strong-coupling regime.  Although the quality of our fit in {\I} does not appear to favour the presence of low-lying excitations, we were not able to accurately measure to sufficiently low temperatures in {\T} to conclusively rule out the existence of $d$-wave superconductivity in this material.  However, it is clear that {\T} has a significantly less conventional superconducting ground state than {\I}, an astonishing difference considering the close similarity between the compounds.  

\begin{table}[h!]
\caption{\label{t:param} Measured and derived anisotropic superconducting parameters in {\T} and {\I}}
\begin{ruledtabular}
\begin{tabular}{lcccc}
{} & \multicolumn{2}{c}{{\T}} & \multicolumn{2}{c}{{\I}} \\
{} & $\parallel$ & $\perp$ & $\parallel$ & $\perp$ \\
\hline
{} & \multicolumn{4}{c}{$Measured$} \\
$T_{c}$ & \multicolumn{2}{c}{4.2~K} & \multicolumn{2}{c}{2.85~K} \\
$H_{c2}(0)$ & 5.9~T & 0.47~T & 4.35~T & 0.25~T \\
$H_{c}(0)$ & \multicolumn{2}{c}{0.0207~T} & \multicolumn{2}{c}{0.0119~T} \\
{} & \multicolumn{4}{c}{$Derived$} \\
$\epsilon$ & \multicolumn{2}{c}{12.6} & \multicolumn{2}{c}{17.2} \\
$\xi$$(0)$ & 940~{\AA} & 75~{\AA} & 1500~{\AA} & 87~{\AA} \\
$\lambda$$(0)$ & 0.12~$\mu$m & 1.5~$\mu$m & 0.13~$\mu$m & 2.2~$\mu$m \\
$\kappa$ & 202 & 1.3 & 260 & 0.87 \\
$G_{3D}$ & \multicolumn{2}{c}{3.3~10$^{-6}$} & \multicolumn{2}{c}{3.0 10$^{-6}$} \\
$G_{1D}$ & \multicolumn{2}{c}{0.36} & \multicolumn{2}{c}{0.69} \\
$\gamma_{n}$ & \multicolumn{2}{c}{0.13~mJ K$^{-2}$ gat$^{-1}$} & \multicolumn{2}{c}{0.13~mJ K$^{-2}$ gat$^{-1}$} \\
BCS gap $\Delta_{0}$ & \multicolumn{2}{c}{$\geq$ 0.9~meV} & \multicolumn{2}{c}{0.4~meV} \\
2$\Delta_{0}/k_{B}T_{c}$ & \multicolumn{2}{c}{$\geq$ 5} & \multicolumn{2}{c}{3.5} \\
$H_{P}$ = $\Delta_{0}/\sqrt{2} \mu_{B}$ & \multicolumn{2}{c}{11~T} &  \multicolumn{2}{c}{4.9T} \\
\end{tabular}
\end{ruledtabular}
\end{table}

Completing the analysis of our data using anisotropic Ginzburg-Landau theory,~\cite{Waldram} we summarise the superconducting parameters of both {\T} and {\I} in Table~\ref{t:param}.  The large calculated values for $\kappa$ highlight both the extreme type II nature of these superconductors and their enormous anisotropy.  $H_{c2}^{\parallel}$ and $\xi^{\parallel}$ should be regarded as minima, due to the high sensitivity of these materials to the field orientation.~\cite{Armici-1980}  We estimate a sample alignment better than $\pm$ 2~$^{\circ}$ with the field; however non-parallel crystalline intergrowths may exist within a single needle-like sample which would reduce our measured $H_{c2}^{\parallel}$.  The measured $H_{c2}^{\parallel}$ = 4.35~T in {\I} approaches the Clogston limit $H_{P}$ = 4.9~T and it would be instructive to re-measure the resistivity of a small single crystal in a high-accuracy goniometer in order to verify the possibility of Pauli-limited superconductivity occurring in this compound.  

\subsection{Normal-State Specific Heat and Phonon Densities of States}

Two features are immediately apparent in a plot of $C_{tot}/T$ vs. $T$: a strong peak at $T {\sim}$ 80~K and a shoulder at $T {\sim}$ 20~K, indicating two dominant phonon energies.  

\begin{figure}[htbp]
\centering
\includegraphics [width=8.5cm,clip] {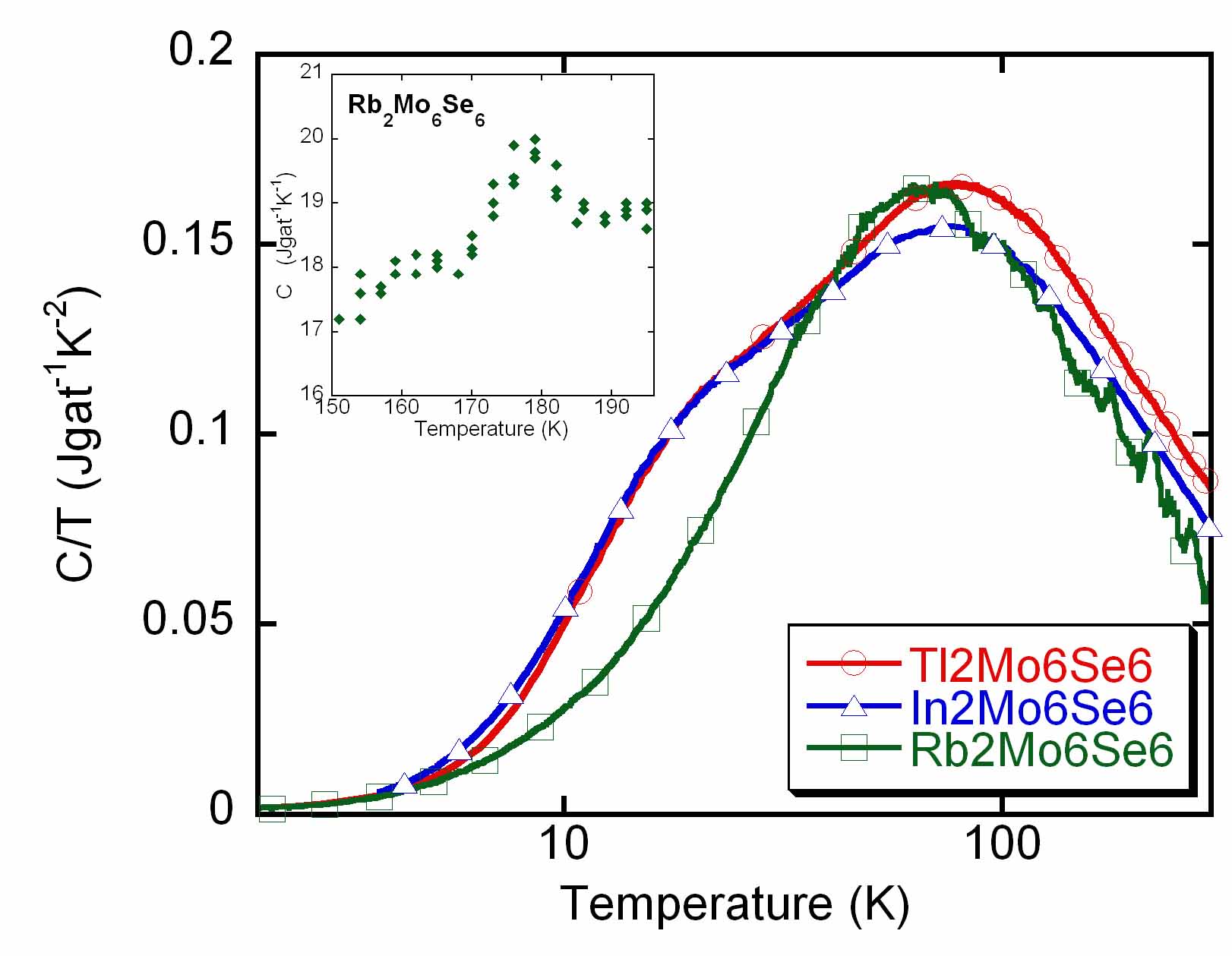}
\caption{\label{Fig_5}Total specific heat divided by temperature $C/T$ for {\T} (circles), {\I} (squares) and {\R} (diamonds).  Note that the increased noise level above 100~K in {\R} can largely be attributed to a low sample mass.  Inset: Close-up of $C(T)$ in {\R}, showing a small peak between 170~K and 185~K.}
\end{figure}

The high-temperature specific data measured on larger polycrystalline samples are sufficiently noise-free to permit an inversion of the phononic contribution $C_{ph}(T)$ to the total heat capacity, hence obtaining the PDoS $F(\omega)$.  We stress that this method does not provide a detailed PDoS map of the type revealed by neutron scattering, but rather produces a smoothed phonon distribution function $\widetilde{F}(\omega)$.  The specific heat and low-temperature features of the PDoS are accurately reproduced by $\widetilde{F}(\omega)$, but it does not give a high-precision representation of the PDoS at high temperature.  We model $F(\omega)$ as a series of logarithmically-spaced Einstein modes with fixed energies and adjustable weights:

\begin{figure}[ht]
\centering
\includegraphics [width=8.5cm,clip] {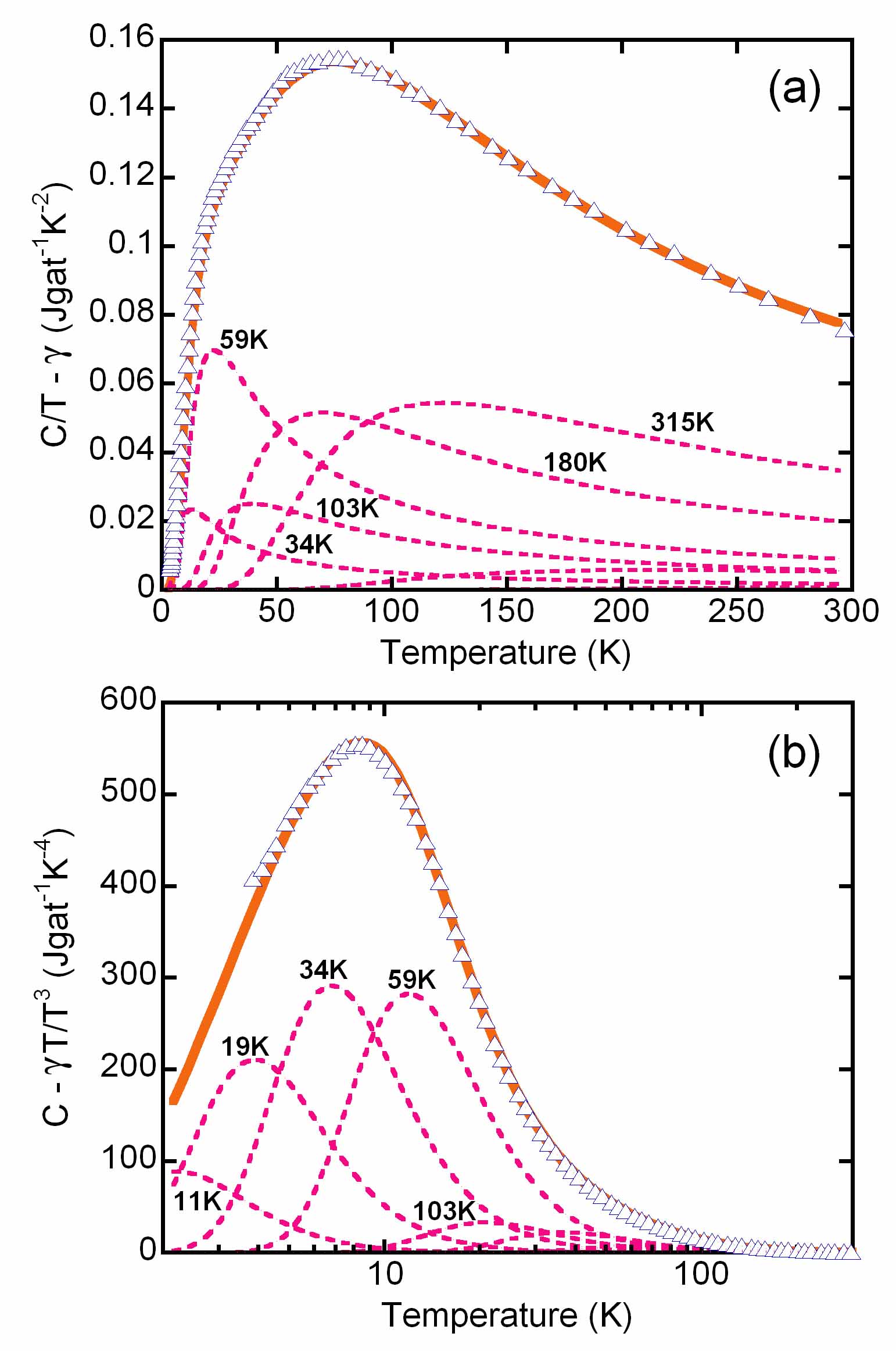}
\caption{\label{Fig_6}(a) Lattice specific heat divided by temperature $C/T-\gamma$ for {\I} (triangles) fitted (solid line) by an assembly of Einstein modes (dashed lines).  (b) Lattice specific heat normalised by temperature cubed $C-{\gamma}T/T^{3}$ (triangles), highlighting the quality of our low temperature fit (solid and dashed lines) and the dominant contribution from the optical mode generated by the In ion.}
\end{figure}

\begin{equation}
F(\omega) = \sum_{k}F_{k}\delta(\omega-\omega_{k})
\end{equation}

Using this representation, the lattice specific heat is then given by
\begin{equation}
C(T) = 3R\sum_{k}F_{k}\frac{x_{k}^{2}e^{x_{k}}}{(e^{x_{k}}-1)^{2}}
\end{equation}

where $x_{k}=\omega_{k}/T$ and $\omega_{k+1}/\omega_{k}=1.75$ to limit the number of modes and provide a stable solution.  A least-squares fit of the measured specific heat for each compound was performed and the decomposition into Einstein modes shown for {\I} as an example in Fig.~\ref{Fig_6}.  The fitting technique for {\T} and {\R} is identical and the results of a similarly high quality, accurately reproducing our experimental data over the entire temperature range.

\begin{figure}[htbp]
\centering
\includegraphics [width=8.5cm,clip] {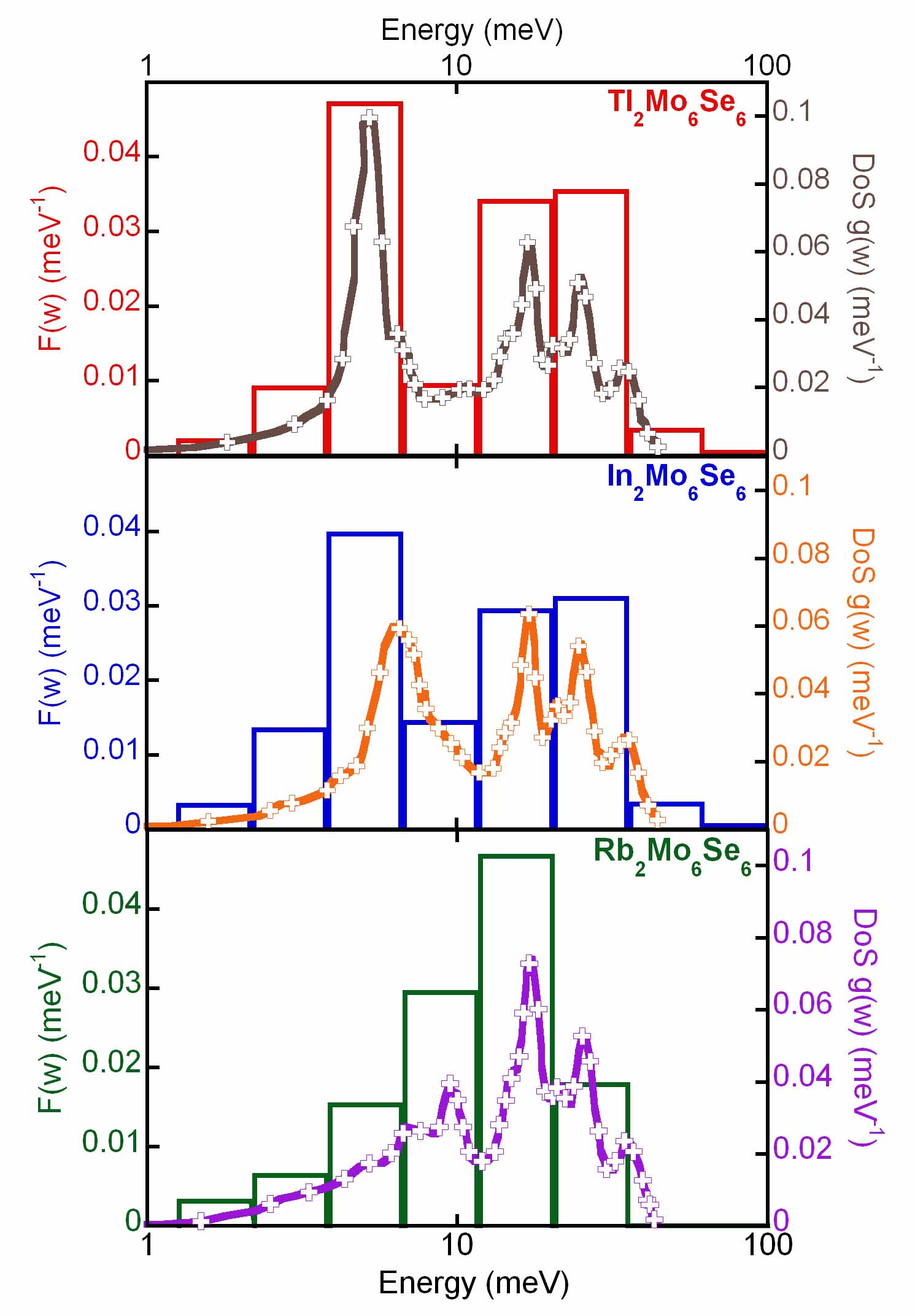}
\caption{\label{Fig_7}Measured phonon density of states for {\T}, {\I} and {\R} (histograms) plotted simultaneously with PDoS data from neutron scattering (crosses).}
\end{figure}

The fitted PDoS are shown for each compound in Fig.~\ref{Fig_7} and the results obtained compare very favourably with previous neutron scattering data.~\cite{Brusetti-1990}  All three compounds clearly display the two key features already identified in the $C/T$ plots: a strong narrow peak below $\sim$ 10~meV and a broader distribution of phonons from $\sim$ 10-30~meV.  The low-energy peak corresponds to an optical mode - the ``guest'' ion phonon - which is mediated by the $M$ ions vibrating in the tubes formed between the Mo$_6$Se$_6$ chains.  The broader maximum at higher energy is mainly due to acoustic intra-chain phonons, including the two Peierls modes discussed earlier.  

Comparing the PDoS of each compound measured, the two superconductors are very similar with a slight spectral weight shift to higher energy in the optical cation mode for {\I}.  This can be explained by considering the smaller atomic mass of In compared to Tl.  However, the intensity of the optical mode in {\R} is significantly reduced and its frequency has been shifted to $\sim$8~meV, compared with $\sim$4-6~meV in {\T} and {\I}.  Neutron scattering data also showed extensive hybridization of the cation mode with the chain modes in {\R}, thus inducing a deformation in the PDoS from 10-30~meV which can also be seen in our data.  

\subsection{Normal-State Resistivity and Electron-Phonon Coupling}

\begin{figure}[ht]
\centering
\includegraphics [width=8.5cm,clip] {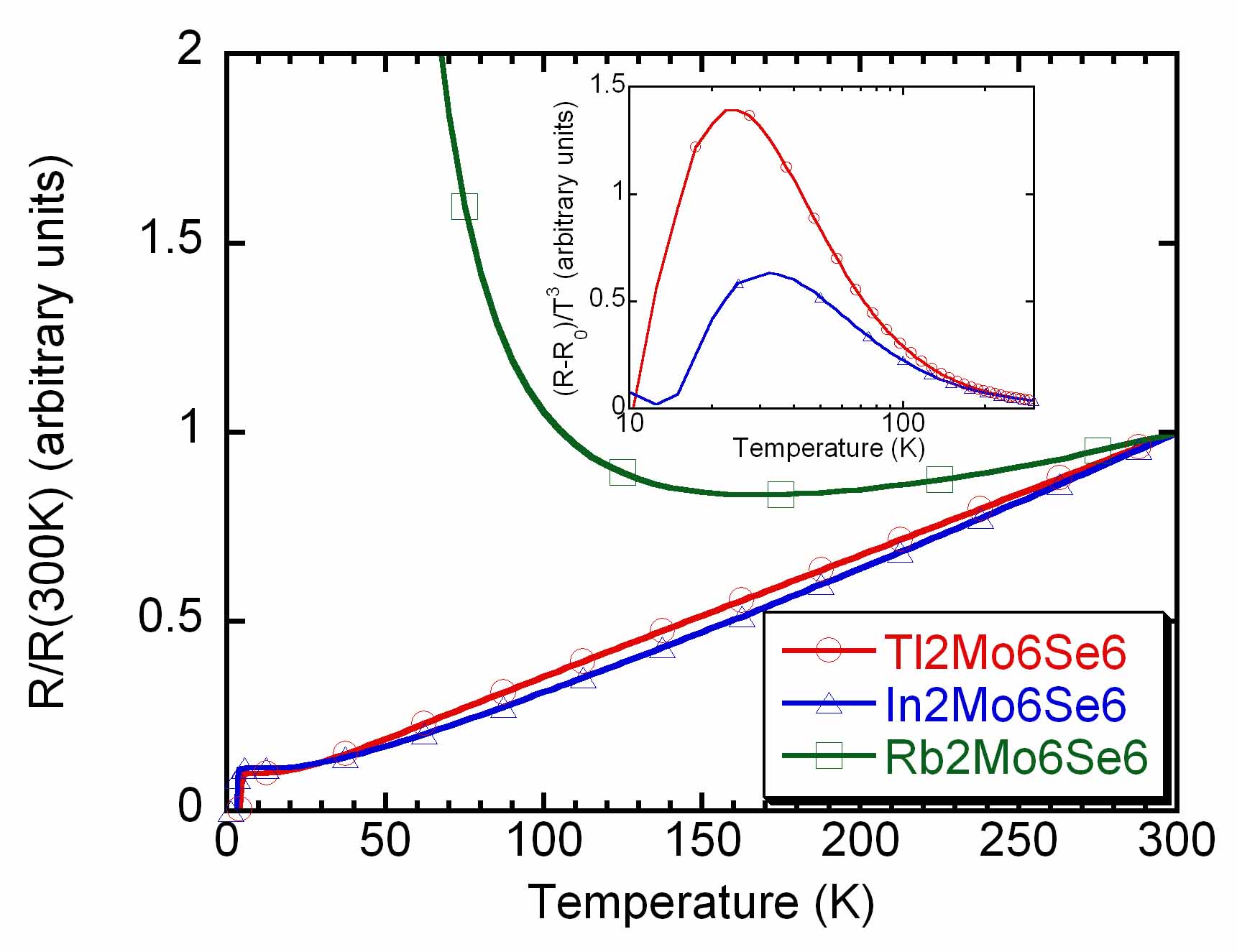}
\caption{\label{Fig_8}Resistivity in {\MMoSe} normalised to 300K. Inset: low-temperature resistivity in {\T} and {\I} normalised to $T^{3}$ after residual subtraction, highlighting the large contribution from low-energy phonons in {\T}.}
\end{figure}

The resistivity curves from 0.35-300~K for each compound are shown in Fig.~\ref{Fig_8}.  {\T} and {\I} both show linear metallic behaviour in the normal state.  There is no evidence for any negative curvature resulting from strong correlation effects or the existence of a Luttinger Liquid ground state at high temperature.~\cite{Giamarchi-1991}  A saturation in the resistivity is observed for $T <$ 15~K in {\T} and {\I}, with residual resistivity ratios of 10.1 and 8.8 respectively.  Conversely, {\R} undergoes a broad metal-insulator transition with a minimum at $T_{c}$ = 170~K and an activation energy $E_{A}$ = 173K.  

\begin{figure}[h!]
\centering
\includegraphics [width=8.5cm,clip] {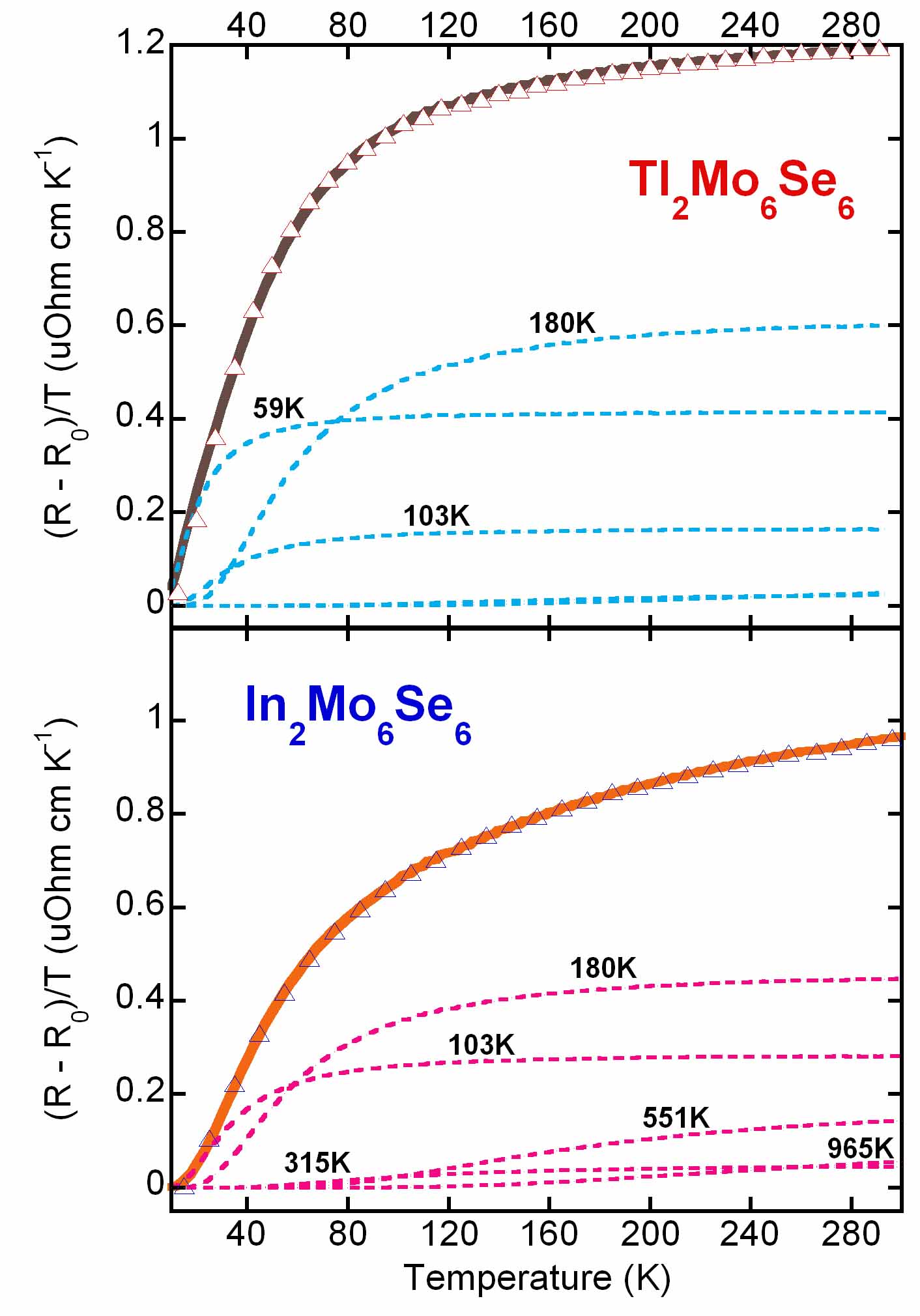}
\caption{\label{Fig_9}Resistivity (with residual subtracted) normalised by $T$ and fitted with the same Einstein mode energies used for the PDoS determination from our specific heat data.}
\end{figure}

Using the same phonon frequency bins as those used to calculate the PDoS, we may now evaluate the electron-phonon coupling from the normal-state resistivity data for {\T} and {\I} (it is not possible to perform this analysis on {\R}, due to the metal-insulator transition).  This procedure has been successfully used to obtain the electron-phonon coupling in the superconducting borides ZrB$_{12}$,~\cite{Lortz-2005} YB$_{6}$~\cite{Lortz-2006} and LuB$_{12}$~\cite{Teyssier-2008} as well as the clathrate superconductors Ba$_8$Si$_{46}$ and Ba$_{24}$Si$_{100}$;~\cite{Lortz-2008} a more detailed account of the method can be found in the references.  Our departure point is the generalised Bloch-Gr\"{u}neisen formula:~\cite{Grimvall}

\begin{figure}[t]
\centering
\includegraphics [width=8.5cm,clip] {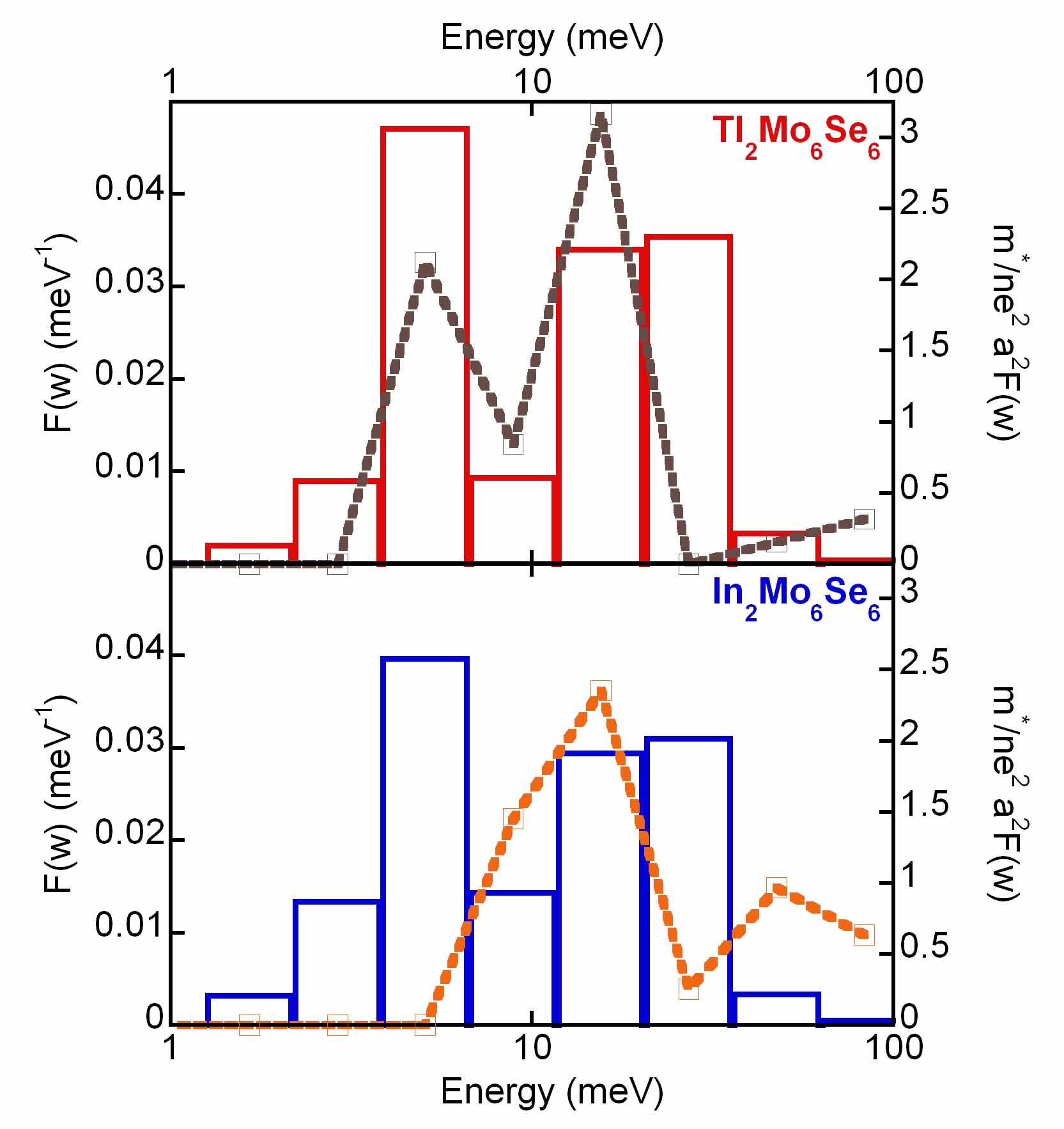}
\caption{\label{Fig_10}Phonon density of states (histogram) and normalised electron-phonon transport coupling function $\alpha_{tr}^{2}F_{\omega}$ (line) for {\T} and {\I}.}
\end{figure}

\begin{equation}
\rho_{BG}(T) = \rho(0) + \frac{4{\pi}m^{*}}{ne^{2}}\int_{0}^{\omega_{max}}\alpha_{tr}^{2}F_{\omega}\frac{xe^{x}}{(e^{x}-1)^{2}d\omega}
\label{BG}
\end{equation}

where $x\equiv\omega/T$ and $\alpha_{tr}^{2}F_{\omega}$ is the electron-phonon ``transport coupling function'' which can be decomposed into Einstein modes to give

\begin{equation}
\alpha_{tr}^{2}F_{\omega} = \frac{1}{2}\sum_{k}\lambda_{tr,k}\omega_{k}\delta(\omega-\omega_{k})
\end{equation}

Substituting this back into equation \ref{BG} yields the discrete version of the Bloch-Gr\"{u}neisen equation:

\begin{equation}
\rho_{BG}(T) = \rho(0) + \frac{2\pi}{\epsilon_{0}\Omega_{p}^{2}}\sum_{k}\lambda_{tr,k}\omega_{k}\frac{x_{k}e^{x_{k}}}{(e^{x_{k}}-1)^{2}}
\end{equation}

where the mode weighting parameters are the dimensionless constants $\lambda_{tr,k}$.  The residual resistivity $\rho(0)$ is determined separately from the raw data, equalling 39.5~$\mu\Omega$~cm for {\T} and 37.2~$\mu\Omega$~cm for {\I}.  

Our fits are shown in Fig.~\ref{Fig_9} and display a clear difference between {\T} and {\I}.  The initial slope of $(R-R_{0})/T$ is much steeper in {\T}, due to a large contribution from a mode with energy 59~K.  In contrast, the lowest energy mode contributing to the resistivity in {\I} is centred at 103~K, with the first significant contribution only arriving at 180~K.  In the absence of any data in the literature for the carrier density $n$, we decompose the unscreened plasma frequency $\Omega_{p}^{2}= ne^{2}/\epsilon_{0}m^{*}$ and express our fitted values for $\alpha_{tr}^{2}F_{\omega}$ normalised by $ne^{2}/m^{*}$.  These are displayed in Fig.~\ref{Fig_10}, superimposed on the PDoS. 

The principal electron-phonon coupling for each compound occurs in the 10-18~meV frequency window.  However, {\T} also exhibits a major coupling in the 3.5-6~meV region, in direct contrast with {\I}, which shows no coupling at all below 6~meV.  $\alpha_{tr}^{2}F_{\omega}$ is intimately related to $\alpha^{2}F_{\omega}$, the electron-phonon coupling function governing superconductivity,~\cite{Allen-1975} implying an additional electronic coupling to the low-energy guest ion phonon in {\T}.

\section{DISCUSSION}
\subsection{Superconducting Transitions}

{\T} and {\I} both display an anomalous broadening of their resistive superconducting transitions, whose amplitude and variation under applied field is not compatible with conventional 3D thermal fluctuation models.  This suggests that the quasi-1D nature of these compounds has a significant influence on the size of the critical region around $T_{c}$.  In 3D systems, the broadening of a superconducting transition under applied magnetic field is due to a finite size effect:~\cite{Lortz-2003} the vortex-vortex separation limits the divergence of the correlation length at $T_{c}$, hence reducing the coherence volume and increasing the importance of thermal fluctuations.  In contrast, a perfect 1D system cannot undergo a phase transition due to insufficient degrees of freedom.  {\T} and {\I} lie in the crossover regime between these two extremes.  

Developing a theoretical model for this transition region is a difficult task; however, we may consider the Peierls transition as an analogous crossover from a quasi-1D system to a quasi-3D ordered state.  Theory predicts a suppression of $T_{c}$ by a factor of up to 4, together with a light smearing of the transition~\cite{Lee-1973} and certain Peierls systems indeed exhibit significantly broadened ``jumps'' in their resistivity as a result of quasi-1D fluctuations.~\cite{Ong-1977}  It therefore seems reasonable to attribute the broadening seen in {\T} and {\I} to the extreme low-dimensional nature of the compound.  

Explaining why the effect is so much more noticeable in {\T} than {\I}, particularly in the specific heat jump, is rather harder especially given that {\I} is more anisotropic.  The calculated 3D Ginzburg number $G_{3D}$ for {\T} is only 10\% larger than that of {\I} and, in any case, $G_{3D}T_{c}$ is several orders of magnitude too small to explain the observed broadening.  Calculating the 1D Ginzburg parameters $G_{1D}$ from the Mishonov model yields more realistic transition width amplitudes, although the fact that the measured transition width for {\T} is larger than that of {\I} implies that this sample was less intrinsically homogeneous.  

The evolution of $G_{1D}$ with applied magnetic field has not yet been calculated for a superconducting nanowire and we were hence unable to track the broadening of the resistive transition using a low-dimensional model.  However, we expect that a field-induced finite size effect similar to that seen in 3D systems should control the transition widths.  The crucial factor here is the difference in coherence volumes $\xi_{\perp}^{2}\xi_{\parallel}$ between the two superconductors: 5290~nm$^{3}$ for {\T} and 11400~nm$^{3}$ for {\I}.  It is well known that low-dimensional fluctuations play an increasingly important r\^{o}le upon the reduction of coherence length in a material; the smaller coherence volume in {\T} must therefore outweigh its lower anisotropy relative to {\I}.  

It is clear that our understanding of low-dimensional fluctuations at a superconducting transition would greatly benefit from a detailed theoretical analysis.  In particular, the reproducible deformation of the specific heat jump in {\T} (which presumably results from a displacement of states from above $T_{c}$ to the back of the jump) is a remarkable phenomenon and merits further attention.  The shape of the jump is reminiscent of that seen in superconducting carbon nanotube matrices,~\cite{Lortz-2009} suggesting that such behaviour may be intrinsic to weakly-coupled superconducting filamentary networks.  A Berezinski-Kosterlitz-Thouless transition is also thought to occur in such materials.~\cite{Wang-2010}  Within this model, the superconducting transition is broadened due to the appearance of a phase-incoherent intermediate state (consisting of Josephson-coupled quasi-1D superconducting fibres) separating the globally-coherent superconducting ground state at low temperature from the metallic normal state above $T_c$.  {\T} and {\I} would be prime candidates to undergo a similar transition.

We note that despite the strong evidence for quasi-1D fluctuations around the superconducting transitions in {\T} and {\I} (as well as the broad resistivity minimum corresponding to the metal-insulator transition in {\R}), there is no indication of a high-temperature Luttinger Liquid ground state in bulk {\MMoSe}.  Transport experiments on Mo$_6$Se$_6$ nanowires have revealed Luttinger behaviour in their conductance, which vanishes as the wire diameter increases above several tens of nanometers.~\cite{Venkataraman-2006}  However, since we are measuring {\MMoSe} crystals with diameters of the order of several hundred microns (and hence an increase in the number of conduction channels by a factor of 10$^8$), the absence of Luttinger effects is to be expected.  

\subsection{Electron-phonon coupling}

Our BCS $s$-wave fits of the specific heat below $T_{c}$ display conventional weak coupling (2$\Delta_{0}/k_{B}T_{c}$ = 3.4) for {\I} and extremely strong coupling (2$\Delta_{0}/k_{B}T_{c} \geq$ 5) for {\T}.  In fact, {\T} may well have usurped the throne of the $\beta$-pyrochlore KOs$_{2}$O$_{6}$ (2$\Delta_{0}/k_{B}T_{c} \geq$ 5) as the strongest-coupling phonon-mediated superconductor currently known.  It should be noted that abnormally strong coupling (ranging up to 2$\Delta_{0}/k_{B}T_{c} \sim 25$) is a characteristic of several quasi-1D CDW systems such as NbSe$_3$ and (TaSe$_4$)$_2$I,~\cite{Fournel-1986,Purdie-1994} due to the transition temperature being suppressed below its mean-field value.  However, given that {\I} is more anisotropic than {\T}, we do not believe that the strong coupling in {\T} originates from its low dimensionality.  

Deconvolving the normal-state resistivity shows that the predominant common electron-phonon coupling for {\T} and {\I} lies in the 10-18~meV range, implying that superconductivity is mediated by the intra-chain modes which (according to neutron diffraction experiments~\cite{Brusetti-1990}) range from 12-40~meV and peak strongly at 17~meV.  This interpretation is supported by a tentative report of superconductivity under pressure in Mo$_{6}$Se$_{6}$.~\cite{Hor-1985-2}  The additional coupling to the low-energy optical mode in {\T} moves this superconductor into the extreme strong-coupling regime, increases $T_{c}$ by nearly 2~K and reduces the coherence volume $\xi_{\perp}^{2}\xi_{\parallel}$, rendering the superconducting transition more susceptible to broadening through quasi-1D fluctuations.  

Our observation immediately begs the question why {\I} does not enjoy a similar coupling to its optical In$^{+}$ mode.  There are two reasons for this: firstly, consider the variation of the hexagonal lattice parameter $a$ and the Pauling radii $R_{p}$ of the Tl$^{+}$, In$^{+}$ and Rb$^{+}$ monovalent cations.  Values for $a$ measured by X-ray diffraction~\cite{Potel-1980,Mori-1984} in {\T}, {\I} and {\R} are given in Table~\ref{t:latt}, together with standard $R_{p}$ values from the literature.  Each cation is located at $a/\sqrt{3}$ from 3 equidistant Mo$_{6}$Se$_{6}$ chains and, since the chain radius is invariant with respect to the cation, the ratio $\sqrt{3}R_{p}/a$ gives a good measure of the freedom of the respective cations to vibrate in their inter-chain tunnels.  

\begin{table}[h!]
\caption{\label{t:latt} Lattice parameters $a$ and cation radii $R_{p}$ in {\MMoSe}}
\begin{ruledtabular}
\begin{tabular}{lccc}
{} & $a$ ({\AA})& $R_{p}$ ({\AA}) & $\sqrt{3}R_{p}/a$ \\
\hline
{\T} & 8.94 & 1.15 & 0.223 \\
{\I} & 8.85 & 1.04 & 0.204 \\
{\R} & 9.26 & 1.48 & 0.277 \\
\end{tabular}
\end{ruledtabular}
\end{table}

It can clearly be seen that the In$^{+}$ ion is less geometrically constrained than Tl$^{+}$ and that Rb$^{+}$ is at considerably less liberty to vibrate than either of its ``superconducting'' counterparts.  This is evident in the neutron scattering data: as pointed out by Brusetti {\itshape{et al.}}, {\R} displays significant hybridisation of the low-energy Einstein phonon, with the higher-energy internal chain modes corresponding to a $\sim$40\% increase in the $M$ ion force constants.  Upon closer examination, the neutron-imaged PDoS of {\I} has a slightly deeper trough at $\sim$11~meV than {\T}, implying marginally less phonon hybridisation, which is consistent with our estimate above.  We therefore believe that the interchain tunnel diameter in {\I} is simply too large relative to the In$^+$ ion to allow its low-energy phonon to effectively couple to the Mo $d$ electrons at the Fermi level in the chains.  

Secondly, the intrinsic electron-phonon coupling strength $\lambda$ is proportional to 1/$\omega^2$, where $\omega$ is the characteristic phonon frequency.~\cite{Grimvall}  Due to its smaller mass, the In$^+$ mode is shifted to higher energy as can be seen both in our PDoS histograms and the neutron data from Brusetti in Fig.~\ref{Fig_7}.  Using the Tl$^+$~=~5.2~meV and In$^+$~=~6.3~meV mode energies from Brusetti {\itshape{et al.}}~\cite{Brusetti-1990}, we calculate $\omega_{\mathrm{In}}^2$/$\omega_{\mathrm{Tl}}^2$ = 1.47; i.e. the coupling to the Tl$^+$ mode should be nearly 50\% stronger than that to the In$^+$ mode.  To make a very crude comparison, we sum our measured $\alpha^{2}F_{\omega}$ from Fig.~\ref{Fig_10} in the relevant energy range 5.1 - 8.9~meV, obtaining $\Sigma~\alpha^{2}F_{\omega}(\mathrm{Tl})$/$\Sigma~\alpha^{2}F_{\omega}(\mathrm{In})$ = 2.04.  This suggests that the frequency-dependent variation in coupling strength and the geometric constraints on the guest ion mode have a roughly equal importance in determining the coupling in {\MMoSe}. 

Naively, we might expect an enormous electron-phonon coupling and ultra-strongly-coupled superconductivity in {\R} due to its narrow effective tunnel diameter.  However, our electronic structure calculations indicate that it is not the geometric constraints on the $M$ ion in {\MMoSe} which determine its anisotropy, but rather the degree of warping in its Fermi sheets.  This is controlled by the in-plane bandwidth $w$, as detailed in Table~\ref{table0}.  The electropositivity of the $M$ ion - its willingness to donate electrons - is a useful quantity for characterising the behaviour of {\MMoSe}, since it is inversely proportional to $w$.  Group IA metals are much more electropositive than Group III and hence {\R} falls victim to a high temperature insulating instability rather than becoming superconducting at low temperature.  (It should nevertheless be noted that high-pressure measurements have succeeded in partially suppressing the metal-insulator transition in {\R} and simultaneously inducing superconductivity with a maximum $T_c$~$\sim$~4~K.~\cite{Hor-1985-2})  In a similar fashion, In is more electropositive than Tl, thus explaining the increased anisotropy seen in the superconducting state of {\I} compared to that of {\T}.  

\subsection{Metal-Insulator Transitions in {\MMoSe}}

Our LDA calculations have indicated two possible mechanisms for the metal-insulator transition in {\MMoSe}: a Peierls transition resulting in the formation of a dynamical CDW, or a SDW driven by strong Coulomb repulsion.  Let us initially consider the SDW scenario.

SDWs and their associated low-temperature antiferromagnetic order are typically imaged using neutron scattering techniques.  Unfortunately, no such data exist in the literature for any of the {\MMoSe} family.  However, a SDW may also be detected by the temperature dependence of its magnetic susceptibility $\chi(T)$.  In a SDW, $\chi(T)$ generally exhibits paramagnetic behaviour above the transition, but then displays a characteristic maximum at the critical temperature $T_{MI}$ before falling, hence signalling the onset of antiferromagnetism.  Any SDW transition should therefore be clearly visible in our ac susceptibility data (Fig.~\ref{Fig_1}).  Examining this closely, no evidence can be seen for any departure from temperature-invariant weakly-diamagnetic behaviour above the noise threshold of our susceptometer (10$^{-8}$~emu).  In particular, the ac susceptibility of {\R} does not display any anomaly as it passes through $T_{MI}$.  We therefore find no experimental evidence for SDW formation in {\MMoSe}.  

Note that our calculated magnetic coupling constants $\lambda_U$ are greater than the critical couplings $\lambda_c$ for all {\MMoSe}, not just {\R}.  This implies that any SDW transition should be present in all three compounds.  However, no $T_{MI}$ may be identified in our resistivity data for {\T} and {\I}.  Furthermore, it is not realistic to suggest that quasi-1D fluctuations suppress SDW formation in {\T} and {\I} but not {\R}: from our electronic structure calculations, {\R} is much more anisotropic than {\T} and {\I} and should hence be more susceptible to such fluctuations.  A far more probable scenario is that low-dimensional fluctuations prevent the SDW transition from taking place in all three compounds, with the metal-insulator transition in {\R} driven by a separate mechanism.  Our LDA calculations for the electron-phonon coupling point towards this being a dynamical CDW and we will continue our discussion from this perspective.  

The transition seen in the electrical resistivity of {\R} is very broad and continuous compared with the distinct steps seen in other CDW materials such as NbSe$_3$ or TaS$_2$.~\cite{Ong-1977,Harper-1977}  Such ``blurring'' can be explained by a combination of the influence of low-dimensional phase fluctuations and the dynamical nature of an underlying CDW whose order parameter may vary in both time and space.  Tarascon~\cite{Tarascon-1984} and Hor~\cite{Hor-1985} have also performed transport measurements on {\R}, obtaining similar broad transitions with $T_{MI} \sim$ 125~K, 104~K and $E_{A}$ = 87~K, 145~K respectively, which are rather smaller than our values $T_{MI} \sim$ 170~K and $E_{A}$ = 173~K.  We attribute the differences between $T_{MI}$ to variations in sample anisotropy, resulting from slight differences in the stoichiometries.  It is well-known that {\T} is not a perfectly stoichiometric compound,~\cite{Brusetti-1994} with the highest $T_{MI}$ occurring in Tl$_{1.95}$Mo$_6$Se$_6$.  Similar non-stoichiometric behaviour occurs in the rest of the {\MMoSe} family, for example producing occasional batches of {\I} which are non-superconducting.~\cite{Lepetit-1984,Mori-1984}  

No clear evidence has been seen for the metal-insulator transition by any other experimental technique, although we observe a small anomaly in the heat capacity of {\R} between 170~K and 185~K (see Fig.~\ref{Fig_5}), just above $T_{MI}$.  However, the height of the peak scarcely exceeds the noise threshold of our data and higher resolution measurements will be required to verify and quantify this feature.  Given the limited evidence for a thermodynamic phase transition at $T_{MI}$, it is tempting to hold intra-chain defects responsible for the insulating behaviour observed at low temperature, relying on thermally-activated interchain hopping to enable metallic conductivity at high temperature.  However, this simple model cannot be applicable to {\MMoSe} since it is unable to explain the broadband conductance noise indicative of sliding density wave motion.~\cite{Tessema-1991}  Furthermore, without a phase transition to deplete the DoS at the Fermi level {\R} would become superconducting at low temperature (unless the defect density is sufficiently high to enable weak localization).  No trace of superconductivity in {\R} has been seen by any experimental technique at ambient pressure.  

An examination of the behavioural trends in the remaining (Group IA) members of the {\MMoSe} family lends further support to the CDW argument.  Since the electronic anisotropy is proportional to the electropositivity of the $M$ atom, we would expect to see metal-insulator transition temperatures fall as we move up Group IA (i.e. Cs$\rightarrow$Na).  Although no systematic study of the family has been carried out by a single author, Tarascon {\itshape{et al.}}~\cite{Tarascon-1984} found that {\C} has a higher T$_{MI}$ than {\R}, which is in line with this hypothesis.  Hor {\itshape{et al.}}~\cite{Hor-1985} have shown that hydrostatic pressure (leading to an enhanced interchain coupling and reduced anisotropy) suppresses T$_{MI}$ to lower temperatures, implying that $T_{MI}$ simply scales with the area of the Fermi surface available for nesting (as would be expected for a CDW).  Nevertheless, a comprehensive study of all Group IA members of the {\MMoSe} family will clearly be required to validate our proposed dynamical CDW scenario.  It is interesting to note that scanning tunnelling spectroscopy on small bundles of Mo$_{6}$Se$_{6}$ chains has revealed metallic behaviour down to 5~K;~\cite{Venkataraman-1999} however the reduced electronic filling in Mo$_{6}$Se$_{6}$ results in a different electronic structure at the Fermi level with up to three bands occupied.~\cite{Ribeiro-2002}  A simple density wave model is hence unlikely to be applicable in this compound.  

In addition to the obvious metal-insulator transition in {\R}, it is important to address the possibility of superconductivity in {\T} arising from a hidden CDW ground state, as suggested by the Hall effect results from Brusetti {\itshape{et al.}}.~\cite{Brusetti-1994}  Our results confirm that there is no discontinuity in the resistivity or jump in the specific heat of either {\T} or {\I} to support this conjecture.  Local stoichiometric heterogeneity in the samples measured could produce a continuous series of local transitions, further blurred by quasi-1D fluctuations.  However, this would produce a positive curvature in $\rho(T)$ for $T <$ 80~K, which is not seen.  Alternatively, if we assume that the inter-chain coupling in {\T} is sufficiently strong for such low-dimensional fluctuations to be negligible, a theoretical model~\cite{Zhou-1988} may be invoked which defines two separate transition temperatures: $T_{cu}$ at which the CDW distortion occurs and $T_{cl}$ at which an energy gap opens over the entire Fermi surface and a metal-insulator transition occurs.  Despite the presence of a structural modulation, metallic behaviour persists in the intermediate temperature range $T_{cl} < T < T_{cu}$ whose width is acutely dependent on the anisotropy.  Using this model and the Brusetti Hall coefficient data, we could identify T $\sim$ 80~K as $T_{cu}$ in {\T} and assume that the superconducting transition takes place at $T_{c} > T_{cl}$.  However, we cannot justify disregarding quasi-1D fluctuations (as required by the model) when they are manifested so clearly in the deformation of the specific heat jump below $T_{c}$ in {\T}.  

A more attractive explanation envisages a partial CDW gradually depleting the Fermi surface in {\T} as the temperature decreases.  This model agrees perfectly with the observed crossover from electron-like to hole-like carriers, yet a corresponding increase in carrier mobility below 80~K is still required to explain the linear metallic resistivity.  It is not clear what type of physical mechanism could be responsible for such a rise in mobility and hence the reported Hall coefficient data in {\T} remains mysterious.  In contrast, the situation in {\R} is less complex: the observed transition to insulating behaviour implies that a gap opens across the entire Fermi surface.  However, with the present data set it is unfortunately impossible to judge whether a similar ``partial CDW'' might be present for $T >$ 170~K in this material.  

Recent work~\cite{Dora-2006} on (TaSe$_4$)$_2$I has suggested the existence of so-called unconventional CDWs (UCDWs) in quasi-1D systems, with a $k$-dependent gap which averages to zero at the Fermi surface.  This results in the formation of a pseudogap prior to a density wave transition and is supported by photoemission data in K$_{0.3}$MoO$_{3}$ and (TaSe$_{4}$)$_{2}$I.~\cite{Dardel-1991}  Within the pseudogap phase, the UCDW competes with low-dimensional quantum fluctuations as described by the Lee-Rice-Anderson model:~\cite{Lee-1973} these act to suppress the Peierls transition temperature $T_P$ well below its mean-field value $T_{MF}$, with the CDW energy gap (the order parameter) fluctuating in time and space between these temperatures.  It seems likely that {\MMoSe} displays a non-trivial combination of low-dimensional fluctuation effects and a dynamical CDW with possible momentum dependence.  Scanning tunnelling spectroscopy (STS) or angle-resolved photoemission spectroscopy (ARPES) would be ideal tools to verify the existence of an UCDW, fluctuating order parameter or dynamical CDW since although a pseudogap should open in the density of states, no static modulation of the structure or charge density is expected.  

\section{CONCLUSIONS}

We have calculated the electronic structures and studied the resistivity and specific heat in both superconducting and normal states for the quasi-1D {\MMoSe} family of cluster condensates ($M$=Tl, In, Rb).  These materials lie on the border between superconducting and insulating instabilities.  
 
Superconductivity in {\T} and {\I} is principally mediated by an internal phonon from the Mo$_6$Se$_6$ chains with an energy in the 12-18~meV range.  It is hoped that theoretical calculations will soon identify the precise energy and nature of this phonon.  {\T} exhibits a further coupling to an optical mode at $\sim$5~meV, which we identify with interchain vibrations of the Tl$^+$ ion.  This extra coupling is not present in {\I} due to the higher In$^+$ phonon energy and the reduced geometrical constraints on the In$^+$ ion between the Mo$_6$Se$_6$ chains.  The low-temperature specific heat of {\I} is well-fitted by a standard single-gap isotropic $s$-wave BCS model with 2$\Delta_{0}/k_{B}T_{c}$ = 3.5$\pm$0.1.  However, the additional interaction with the low-energy mode in {\T} pushes it into the strong coupling r\'{e}gime with 2$\Delta_{0}/k_{B}T_{c}~{\geq}~5$.  The specific heat jump at $T_{c}$ also exhibits a significant deformation with a shift in states to low temperature, which we attribute to strong low-dimensional fluctuations accentuated by the small coherence volume.  STS or similar tunnelling experiments would provide conclusive proof of the gap symmetry, as well as displaying the coupling to the low-energy phonon.  

Our LDA calculations show that in all members of the {\MMoSe} family, a single 1D helix band crosses the Fermi level.  Its in-plane dispersion $w$ is reduced by a factor of ten in {\R} compared with {\T} and {\I}, while the out-of-plane dispersion is practically unchanged.  Using an analytical model, we have shown that this reduction of $w$ is sufficient to explain the trend from metallic conductivity followed by strongly-coupled superconductivity in {\T} to a high-temperature metal-insulator transition in {\R}.  This insulating ground state is a consequence of either a dynamical Peierls (CDW) instability or a SDW transition driven by strong electronic correlations. The temperature invariant magnetic susceptibility seen in all three compounds favours a CDW interpretation.  {\MMoSe} are therefore an ideal target for future ARPES experiments, since a dynamical CDW should generate a pseudogap at the Fermi level.

\section{ACKNOWLEDGEMENTS}

We thank A. Junod, M. Hoesch, Y. Fasano, C. Bernhard and L. Forr{\'o} for invaluable advice and discussions. This work was supported by the National Centre of Competence in Research MaNEP and the Swiss National Science Foundation.

\end{document}